\documentclass{jfm}

\usepackage{epsfig,color,soul}
\usepackage[dvipsnames]{xcolor}
\usepackage{float}
\usepackage{natbib}
\usepackage{amsmath,amssymb,graphicx,bm,gensymb} 
\usepackage{mathptmx} 
\usepackage[outdir=./]{epstopdf}
\usepackage[utf8]{inputenc}

\usepackage{array}
\newcolumntype{L}[1]{>{\raggedright\let\newline\\\arraybackslash\hspace{0pt}}m{#1}}
\newcolumntype{C}[1]{>{\centering\let\newline\\\arraybackslash\hspace{0pt}}m{#1}}
\newcolumntype{R}[1]{>{\raggedleft\let\newline\\\arraybackslash\hspace{0pt}}m{#1}}

\def\ba#1\ea{\begin{align}#1\end{align}}
\def\bsa#1#2\esa{\begin{subequations}\label{#1} \begin{align}#2\end{align} \end{subequations}}

\def\lp{\left(}\def\rp{\right)}\def\lb{\left[}\def\rb{\right]}
\def\lcb{\left\{}\def\rcb{\right\}}
\allowdisplaybreaks

\definecolor{darkblue}{RGB}{83,0,93}

\def\NA{\bm{\nabla}}\def\f{\frac}\def\p{\partial}

\def\u{{\bf{u}}}

\def\ba#1\ea{\begin{align}#1\end{align}}
\def\bsa#1#2\esa{\begin{subequations}\label{#1}
\begin{align}#2\end{align} \end{subequations}}

\title[]{Topography generation by melting and freezing in a turbulent shear flow}

\author[]{Louis-Alexandre Couston$^{1,2,3}$\thanks{loton@bas.ac.uk}, Eric Hester$^4$, Benjamin Favier$^5$, John R. Taylor$^2$, Paul R. Holland$^1$, and Adrian Jenkins$^{1,6}$} 

\affiliation{$^1$ British Antarctic Survey, Cambridge, CB3 0ET, UK \\  $^2$ Department of Applied Mathematics and Theoretical Physics, University of Cambridge, Cambridge, UK \\ $^3$ Univ Lyon, Ens de Lyon, Univ Claude Bernard, CNRS, Laboratoire de Physique, F-69342 Lyon, France \\ $^4$ Department of Mathematics and Statistics, University of Sydney, Australia \\ $^5$ Aix Marseille Univ, CNRS, Centrale Marseille, IRPHE, Marseille, France \\ $^6$ Department of Geography and Environmental Sciences, Northumbria University, Newcastle upon Tyne, UK}

\begin{document}
\maketitle

\begin{abstract}

{We report an idealized numerical study of a melting and freezing solid adjacent to a turbulent, buoyancy-affected shear flow, in order to improve our understanding of topography generation by phase changes in the environment}. We use the phase-field method to dynamically couple the heat equation for the solid with the Navier-Stokes equations for the fluid. We investigate the evolution of an initially-flat and horizontal solid boundary overlying a pressure-driven turbulent flow. We assume a linear equation of state for the fluid and change the sign of the thermal expansion coefficient, such that the background density stratification is either stable, neutral or unstable. We find that channels aligned with the direction of the mean flow are generated spontaneously by phase changes at the fluid-solid interface. Streamwise vortices in the fluid, the interface topography and the temperature field in the solid influence each other and adjust until a statistical steady state is obtained. The crest-to-trough amplitude of the channels are larger than about 10$\delta_{\nu}$ in all cases, with $\delta_{\nu}$ the viscous length scale, but are much larger and more persistent for an unstable stratification than for a neutral or stable stratification. This happens because a stable stratification makes {the cool melt fluid} buoyant such that it shields the channel from further melting, whereas an unstable stratification makes {the cool melt fluid} sink, inducing further melting by rising hot plumes. The statistics of flow velocities and melt rates are investigated, and we find that channels and keels emerging in our simulations do not significantly change the mean drag coefficient.
\end{abstract}

\section{Introduction}\label{sec:intro}

Melting and freezing processes between ice and water play an important role in the environment. For instance, the melting of ice shelves, i.e., floating glacial ice, can lead to reduced buttressing of the grounded polar ice sheets and increased sea-level rise \cite[][]{Pritchard2012,Rignot2013,Kennicutt2019}, while freezing of high-latitude oceans by a cold atmosphere results in sea-ice formation, increased albedo and increased ocean salinity through brine rejection {\cite[][]{Wells2019}}. Icebergs, ice shelves and sea ice are km-scale objects with long lifetimes but their evolution is controlled by heat and salt fluxes across mm-thin ice-water boundary layers, which fluctuate rapidly \cite[][]{Dinniman2016}. The front of a marine-terminating glacier can melt as fast as several meters per day horizontally \cite[as recently reported for LeConte Glacier,][]{Sutherland2019}, but an ice shelf around Antarctica typically melts at a rate of only a few centimeters per day or less \cite[][]{Dutrieux2014}. On the other hand, ocean currents are most often turbulent and exhibit temporal variabilities down to just a few seconds \cite[][]{Davis2019}, such that phase changes between ice and water are multi-physics phenomena with large scales separation. 

An important consequence of phase changes is that topographical features can emerge at the ice-water interface when the rate of melting and freezing is spatially variable. Basal channels \cite[][]{Stanton2013,Gourmelen2017} and terraces \cite[][]{Dutrieux2014} have been observed at hundreds-of-metre to kilometre scales under ice shelves, the underside of icebergs exhibits ablation channels at metre scale and scallops at {tens-of-centimetre scale} \cite[][]{Hobson2011}, and, more generally, rough features can be seen  from centimetre scale to tens-of-meter scale under sea ice \cite[][]{Wadhams2006,McPhee2008,Lucieer2016} and up to kilometre scale under ice shelves \cite[][]{Nicholls2006}. The interplay between flow dynamics and phase changes leading to the generation and persistence of topographical features in the environment is of fundamental importance. Indeed, the presence of topography can significantly affect the long-term flow dynamics as well as the average melting or freezing rate of the ice boundary, as suggested by, e.g., the large spatial variability of melting of basal terraces \cite[][]{Dutrieux2014} and recent laboratory experiments on ice scallops \cite[][]{Bushuk2019}. 

{Buoyancy forces play an important role in the coupling between phase changes, flow dynamics and topography generation. Buoyancy forces can be stabilizing or destabilizing depending on the relative orientation between the gravitational acceleration and the density gradient. In polar seas, cold and fresh melt water near the ice boundary is lighter than the surrounding, such that buoyancy forces are restoring below a horizontal ice boundary (e.g., below an ice shelf) and drive upwellings along a vertical ice face (e.g., at the front of a marine-terminating glacier). In a cold freshwater system, the thermal expansion coefficient of water is negative (it is negative for temperatures $0^\circ$C $<T<4^\circ$C at atmospheric pressure; cf. \cite[][]{Thoma2010a}), such that buoyancy forces are stabilizing for water under an ice cover (e.g., as in a frozen lake) but destabilizing for water above ice (e.g., for a supraglacial lake or river).}

For a system dominated by destabilizing buoyancy forces, the interplay is strong between fluid dynamics and phase topography. In a thermally-stratified fluid with finite depth below a solid phase, the unstable density stratification sets up a large-scale circulation known as Rayleigh-Bénard convection with alternating warm upwelling and cold downwelling regions. The warm upwellings drive stronger melting than the cold downwellings, such that a topography can emerge from spatially-variable heat fluxes. The topography enhances the large-scale circulation, such that a positive feedback is obtained and the flow dynamics and solid boundary become phase locked \cite[][]{Rabbanipour2018,Favier2019}. {Dissolution of a phase boundary, i.e., with phase changes driven by concentration gradients rather than temperature effects, in a gravitationally-unstable fluid can also lead to convective motions and the generation of three-dimensional topography (even in the absence of a large-scale circulation), as shown by experiments \cite[][]{Kerr1994,Sullivan1996} and numerical simulations \cite[][]{Philippi2019}. Streamwise patterns also emerge in dissolution experiments when the phase boundary is not perpendicular to gravity but inclined \cite[][]{Allen1971,Cohen2020}.}
 
{Despite the existence of many studies on pressure-driven and shear flows \cite[][]{Kelly1994,Zonta2018}, the possibility for topography to emerge between a horizontal boundary-layer flow and a solid phase, i.e., perpendicular to gravity, is not well understood, at least compared to the case of topography generation by Rayleigh-Bénard convection}. Boundary layer flows strongly affected by shear, such as under ice shelves, are yet as common (if not more) as buoyancy-driven flows in the environment, such that there is intense interest in predicting their ability to generate topographical features (or roughness) at horizontal ice boundaries and the impact the sustained topography can have on overall melt rates. Using laboratory experiments, \cite{Gilpin1980} demonstrated the existence of an interfacial instability and the generation of ripples at an ice boundary below a horizontal turbulent boundary-layer water flow. The experiments had an unstable density stratification owing to the negative thermal expansion coefficient of freshwater at low temperatures \cite[][]{Toppaladoddi2019}, but still demonstrated that shear does not always prevent topography generation. {The necessary condition for an interfacial instability to develop, regardless of the type of density stratification, is that the maximum of mass transfer from the solid to the fluid due to a heat flux or concentration gradient (resulting in ablation) at the boundary be shifted by $-90^\circ$ to $+90^\circ$ compared to the maximum (crest) of boundary topography \cite[][]{Hanratty1981}.} Recently, \cite{Claudin2017} demonstrated that such a shift is possible for a horizontal neutral turbulent flow and proposed a saturation mechanism for the finite amplitude of two-dimensional scallops. Three-dimensional effects and buoyancy forces are expected to play an important role on topography generation and melting rates but were not considered in the study of  \cite{Claudin2017}, which also relied on parameterized flow non-linearities. Thus, additional efforts are necessary to improve our understanding of the physical mechanisms leading to the generation and saturation of phase topography by horizontal shear flows.

Here we demonstrate the possibility to use direct numerical simulations to investigate the generation of topography at a phase boundary adjacent to a shear flow affected by buoyancy. We focus on the case of an initially-flat and horizontal solid, i.e., perpendicular to gravity, and investigate the influence of density stratification on the topography obtained and the coupled fluid-solid dynamics. Our numerical models solves the evolution of the fluid and solid phases simultaneously using the phase-field method. The phase-field method is a one-domain two-phase fixed-grid method that was originally developed by the metallurgy community for relatively smooth flows \cite[][]{Wang1993,Karma1998,Beckermann1999}, but which was  applied to the case of vigorous convective flows recently \cite[][]{Favier2019,Purseed2020}. Other methods that simultaneously solve for the evolution of a fluid phase and a solid phase include the enthalpy method \cite[][]{Ulvrova2012}, the level set method \cite[][]{Gibou2007}, the lattice-Boltzmann method \cite[][]{Rabbanipour2018} and two-domain moving-boundary methods \cite[][]{Ulvrova2012}. The main advantage of the phase-field method over these other methods is that it can be implemented relatively easily in any fluid solver.

Our study aims to contribute to the physical understanding of  topography generation by shear flows at horizontal boundaries and the associated changes in mean melt rates, as investigated most recently theoretically by \cite{Claudin2017} and experimentally by \cite{Bushuk2019}. {Numerical constraints force us to consider an idealized setup, however, such that our fluid and solid phases are not exactly representative of water and ice. Notably, we assume that the fluid and solid have the same thermodynamical and transport properties, i.e., e.g., same thermal conductivity, and we consider an anomalously warm fluid in order to minimize the time scale separation between the turbulent dynamics and generation of boundary topography. Due to computational constraints, the external flow in our simulations is also weaker than what may be expected for scallops formation \cite[][]{Claudin2017,Bushuk2019}.}

The main result of our paper is that topographical features spontaneously emerge at the ice-water interface due to uneven melting of the solid boundary by the shear flow. We investigate the effect of background density stratification and we demonstrate that the topography is  dominated by keels and channels that are aligned with the direction of the mean flow in all cases. 

We organize the manuscript as follows. In \S\ref{sec:model} we describe the phase-field method, the dimensionless equations and the numerical method. In \S\ref{sec:results} we present and discuss the direct numerical simulation results obtained for three different background stratifications. {In \S\ref{sec:geophysics} we discuss the link between our results and geophysical applications and explain why we did not observe three-dimensional scallops.} In \S\ref{sec:discussion} we conclude. Finally, in appendices \S\ref{sec:appA}-\S\ref{sec:appD}, we provide additional details about the method and results.

\section{Model}\label{sec:model}

\subsection{Phase-field method}

We investigate the generation of topography due to uneven melting and freezing at a fluid-solid interface. The solid is fixed and located above the fluid where a Poiseuille/channel flow develops due to an external pressure gradient (see figure \ref{fig1}). {The initial thickness of the fluid (resp. solid) layer is $H$ (resp. $H/2$), such that the channel full depth is $3H/2$.} The domain length (in the direction of the flow) is $L_x=4\pi H$ and the transverse width is $L_y=2\pi H$. We define a Cartesian coordinate system $(x,y,z)$ centered on the bottom of the channel with $z$-axis vertical upward, i.e., opposite to gravity, and use superscripts $^{(f)}$ and $^{(s)}$ to denote variables in the fluid and the solid, respectively. The fluid velocity $\u^{(f)}$ and pressure $p^{(f)}$ evolve according to the Navier-Stokes equations under the Boussinesq approximation. {For simplicity, we assume that the solid and fluid phases have the same thermodynamical and transport properties, i.e., the same reference density $\rho_f$, the same specific heat capacity at constant pressure $c_p$ and the same thermal conductivity $k$. Thus, the temperatures $T^{(f)}$ and $T^{(s)}$ evolve according to the same advection-diffusion (energy) equation, which turns into the heat equation where the velocity is zero. We consider a generic linear equation of state for the fluid, i.e., not specific to water, such that
\ba{}
\rho^{(f)}=\rho_f\left(1-\alpha T^{(f)}\right),
\ea
with $\alpha$ the thermal expansion coefficient. For a {pure-component} flow, the fluid-solid interface must be at the temperature of melting, denoted by $T_m$, and the movement of the interface is governed by the Stefan condition, i.e.,
\bsa{01}
& T^{(f)}=T^{(s)}=T_m, \\
& \rho_s\mathcal{L}v_n= q_n^{(f)}-q_n^{(s)} = -k_f\hat{\bf{n}}\cdot\NA T^{(f)}+k_s\hat{\bf{n}}\cdot\NA T^{(s)} ,
\esa
where $\rho_s$ is the reference density of the solid, $v_n$ is the interface velocity in the direction normal to the interface and directed toward the solid phase (supported by unit vector $\hat{\bf{n}}$), $\mathcal{L}$ is the latent heat of fusion per unit mass, $q_n$ is the heat flux in direction $\hat{\bf{n}}$, $k_s$ (resp. $k_f$) is the thermal conductivity of the solid (resp. fluid) and $\NA$ is the gradient operator \cite[][]{Worster2000}. We recall that we assume the same properties for the two phases, i.e., such that in our case $k_f=k_s=k$ and $\rho_s=\rho_f$ in equation \eqref{01}. Note that the properties of water and ice are different under typical atmospheric pressure and near-freezing temperature conditions, i.e., such that $\rho_f\approx 999$ kg/m$^3$, $c_{pf}\approx 4200$ J/kg/K and $k_f\approx 0.6$ W/m/K, while $\rho_s\approx 917$ kg/m$^3$, $c_{ps}\approx 2100$ J/kg/K and $k_s\approx 2.2$ W/m/K. The relative differences are small, however, i.e., within a factor of 4 or less, such that we do not expect fundamental differences between our model results and physical processes involving water and ice in nature.} 

Here we use a volume-penalization method \cite[][]{Angot1999}, which is a type of immersed boundary method, combined with the phase-field method, in order to solve for phase-change processes and the evolution of the variables in the fluid and the solid simultaneously. Specifically, we solve the Navier-Stokes equations in the Boussinesq approximation and the {advection-diffusion (energy) equation} for temperature combined with an equation for the fluid fraction $\phi$, i.e.,
\bsa{00}
&\f{\p \u}{\p t} + \phi \lp \u\cdot\nabla \rp \u  = \nu \nabla^2\u - \f{1}{\rho_f}\nabla p + \alpha g T \hat{\bf{z}} + \f{\Pi}{\rho_f}\hat{\bf{x}} - \f{(1-\phi)}{\tau_p}\u,\label{001}\\
&\f{\p T}{\p t} + \phi \lp \u\cdot\nabla \rp T  = \kappa \nabla^2T - \f{\mathcal{L}}{c_p}\f{\p \phi}{\p t},\label{002}\\
& \f{\p \phi}{\p t} = a \nabla^2 \phi + b\phi(1-\phi)\lb 2\phi-1+c(T-T_m)\rb,\label{003} \\
&\nabla\cdot\u=0,\label{004}
\esa
with $\u=(u,v,w)$, $p$, and $T$ defined in both the fluid and solid, i.e., assuming that the fluid and solid phases form a single domain, such that we drop the superscripts $^{(f)}$ and $^{(s)}$. In equations \eqref{00}, $\nu$ is the constant kinematic viscosity, $g$ is the gravity acceleration, $\Pi$ is the imposed pressure-gradient force and $\kappa=k/\rho_f/c_p$ is the constant thermal diffusivity; $\tau_p$, $a$, $b$ and $c$ are parameters related to volume penalization and the phase-field method, which we define later, and $\hat{\bf{z}}$ and $\hat{\bf{x}}$ are the unit vectors of the $z$ and $x$ axis, respectively. Note that the third term on the right-hand-side of equation \eqref{001} represents the buoyancy force.

The fluid fraction, $\phi$, also known as the phase-field variable or order parameter, satisfies a forced diffusion equation \eqref{003} with parameters tuned such that $\phi$ transitions continuously from 1 in the fluid to 0 in the solid, across a diffuse interface whose thickness is artificial and must be smaller than all physical length scales in the problem, {including the viscous length scale} (cf. appendix \ref{sec:appA}). $\phi$ is introduced in the momentum \eqref{001}, energy \eqref{002} and continuity \eqref{003} equations, in order to modulate locally the importance of each physical processes based on the component's phase. For instance, the last term on the right-hand-side of equation \eqref{001} is a linear (penalization) damping term, which is active in the solid but inactive in the fluid, while the second term on the right-hand-side of the energy equation \eqref{002} is a heat sink or source that represents the consumption or release of latent heat associated with melting or freezing.  In the limit of infinitesimally-small diffuse interface thickness of the phase field, it has been shown that the dynamics of the fluid-solid interface governed by equations \eqref{00} converges to the exact Stefan conditions \eqref{01}, and that the fluid velocity converges to 0 at the fluid-solid interface, thus mimicking a no-slip boundary. Here we multiply by $\phi$ the advective terms in equations \eqref{001}-\eqref{002}, such that they are zero in the solid phase. Previous studies have used both damped and undamped advective terms and we discuss the impact of our choice on the results in appendix \ref{sec:appB}. 

\subsection{Dimensionless equations}

Equations \eqref{00} can be non-dimensionalized in order to identify the set of independent control parameters. {Following previous studies \cite[e.g.,][]{Favier2019}, we use the thermal diffusive time scale $\tau_{\kappa}=H^2/\kappa$ as normalizing time scale, i.e., such that the dimensionless variables, denoted by tildes, are defined as
\ba{}\label{02}
(x,y,z)=(H\widetilde{x},H\widetilde{y},H\widetilde{z}), \; t=\tau_{\kappa}\widetilde{t}, \; u=u_{\kappa}\widetilde{u}, \; T=T_m+\Delta T\widetilde{T}, \; p = \rho_f u_{\kappa}^2\widetilde{p}, \; \phi=\widetilde{\phi},
\ea
with $u_{\kappa}=\kappa/H$ the diffusion velocity scale, and $\Delta T=T_b-T_m$ is the temperature scale with $T_b$ the dimensional temperature on the bottom boundary. The time scale $\tau_{\kappa}$ is particularly relevant for discussing the long-term dynamics of the system since temperature evolves in the solid through diffusion. We will use the shorter friction time scale to describe relatively rapid processes, such as convection in the fluid (see section \S\ref{sub:vars}). }

Substituting variables \eqref{02} into equations \eqref{00}, and dropping tildes, we obtain the dimensionless equations
\bsa{03}
&\f{\p \u}{\p t} + \phi \lp \u\cdot\nabla \rp \u  = Pr \nabla^2\u - \nabla p + Pr Ra  T \hat{\bf{z}} + 2Pr^2 Re \hat{\bf{x}} - Pr\f{(1-\phi)}{\Gamma}\u,\label{031}\\
&\f{\p T}{\p t} + \phi \lp \u\cdot\nabla \rp T  = \nabla^2T - St\f{\p \phi}{\p t},\label{032}\\
& \f{\p \phi}{\p t} = A \nabla^2 \phi + B\phi(1-\phi)(2\phi-1+CT),\label{033} \\
&\nabla\cdot\u=0.\label{034}
\esa
{The control parameters in equations \eqref{03} are the Prandtl number, $Pr$, which compares kinematic viscosity to thermal diffusivity, the centreline Reynolds number, $Re$, which compares the pressure gradient force to viscous dissipation, the Rayleigh number, $Ra$, which compares buoyancy forces to viscous and thermal dissipation, and the Stefan number, $St$, which compares the available sensible heat to the latent heat. They are related to the physical parameters through
\ba{}\label{dlesspar}
Pr=\frac{\nu}{\kappa}, \quad Re=\frac{\Pi H^3}{2\rho_f\nu^2}, \quad Ra = \frac{\alpha g \Delta T H^3}{\nu\kappa}, \quad St=\frac{\mathcal{L}}{c_p\Delta T}.
\ea
}The additional parameters $\Gamma=\tau_p\nu H^2/\kappa^2$, $A=a/\kappa$, $B=b/(\kappa/H^2)$ and $C$ are  non-physical and prescribed based on numerical constraints of the volume-penalization and the phase-field methods (cf. appendix \ref{sec:appA}). The problem is fully specified once $Pr$, $Re$, $Ra$ and $St$ are known and the boundary conditions are prescribed. Here we enforce a no-slip, fixed-temperature condition at the top of the ice, i.e., $\u=\bf{0}$ and $T=T_t<0$, at $z=1.5$. We impose free-slip, fixed-temperature conditions on the bottom boundary, i.e., $\p_zu=\p_zv=w=0$ and $T=1$, at $z=0$, such that we simulate only one half of a full channel flow (to reduce computational costs). The dimensionless melting temperature is $T=0$. The initial interface position is $z=1$ and we note $(l_x,l_y,l_z)=(4\pi,2\pi,1.5)$ the domain lengths in dimensionless space. The initial condition in the fluid is a half-channel laminar Poiseuille flow superimposed with divergence-free white noise for the velocity fluctuations. 

We will generally discuss our results in terms of the steady-state friction (or shear) Reynolds number, $Re_*$, and the friction Richardson number, $Ri_*$, i.e.,
\ba{}\label{ndlnum}
Re_*=\sqrt{2Re}=\sqrt{\f{\Pi H^3}{\rho_f\nu^2}}, \quad Ri_*=\f{-Ra}{PrRe}=\f{-2\rho_f\alpha\Delta T g}{\Pi},
\ea
since they are more commonly used than $Re$ and $Ra$ in turbulent channel flow studies \cite[][]{Garcia-Villalba2011,Zonta2018}. {The key difference between $Re$ and $Re_*$ is that the former is based on the velocity on the bottom free-slip boundary of the channel in the laminar regime, which is $[\Pi H^2(1-z^2/H^2)/(2\rho_f\nu)]|_{z=0}$ using dimensional variables, while the latter is based on the friction velocity, which is $\sqrt{-\tau_w}=\sqrt{\Pi H/\rho_f}$ with $\tau_w$ the mean wall shear stress, again using dimensional variables. Here, we favour the friction Richardson number over the Rayleigh number as control parameter, even when the stratification is unstable, because they are both input parameters and because the wall shear stress is an important driver of turbulence in all cases. The importance of shear forces compared to buoyancy forces can be estimated from the Monin-Obukhov length, which is
\ba{}\label{LMO}
L_{MO} = \f{Re_*^3Pr^2}{NuRa},
\ea
in terms of dimensionless variables and which is often reported in mixed-convection experiments \cite[][]{Pirozzoli2017,Blass2020}, with $Nu$ the Nusselt number, which we define later (see equation \eqref{definitions}). The Monin-Obukhov length estimates the distance from the boundary within which shear is as important or more important than buoyancy. In our simulations, we always have $L_{MO}>0.97$, such that shear is a significant source of turbulence throughout the domain.}

\begin{figure}
\centering
\includegraphics[width=1\textwidth]{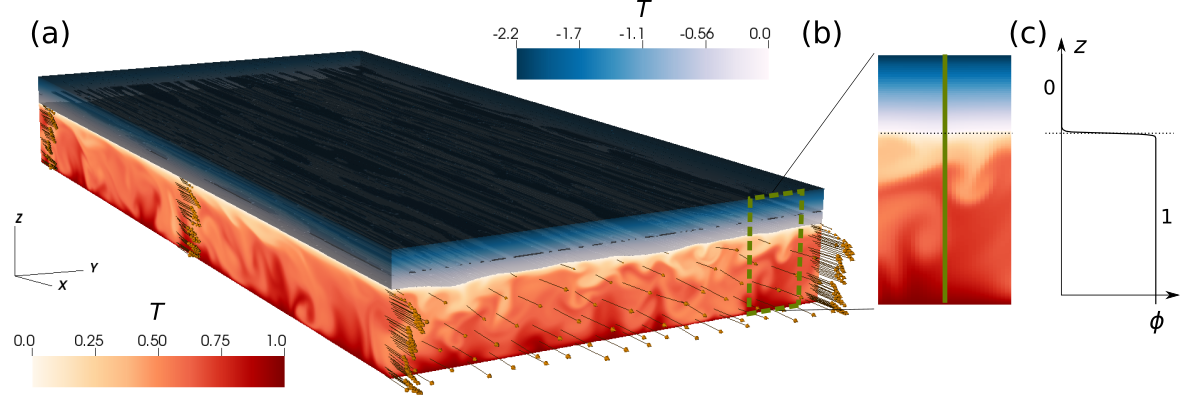}
\vspace{-0.in}\caption{(a) Simulation snapshot showing the temperature field in the fluid (red colormap) and the solid (blue colormap) for neutral stratification ($Ri_*=0$) at late time, i.e, such that it is representative of the statistical steady state. {The arrows along the four vertical transects and across the front face $x=l_x$ (displayed over a limited number of points for clarity) show the velocity vectors. The velocity vectors all point primarily in the direction of the mean flow, $+\hat{x}$.} (b) Zoom-in on a region of (a). (c) Variation of the phase field from $\phi=1$ in the fluid to $\phi=0$ in the solid along the vertical solid line drawn in (b). The non-dimensional lengths in $x$, $y$ and $z$ directions are $l_x=4\pi$, $l_y=2\pi$ and $l_z=1.5$, respectively.}
\label{fig1}
\end{figure}

\begin{table}
\centering
	\begin{tabular}{C{1.6cm}C{0.6cm}C{0.8cm}C{0.8cm}C{0.7cm}C{0.6cm}C{1.0cm}C{0.8cm}C{1.4cm}C{0.9cm}C{1cm}}
stratification & $Ri_*$  & $Re^I_b$ & $Re^{III}_b$ & $Nu^I$ & $q^s$ & $Nu^{III}$ & $\xi^{III}$ & $\lp\f{\xi_-}{\delta_{\nu}},\f{\xi_+}{\delta_{\nu}}\rp$ & $10^3C_D^I$ & $10^3C_D^{III}$  \\ \hline
stable 			& 40 & 2720 & 2630 & 3.02 & 2.83 & 3.09  & 1.042 & (2.4,2.2) & 5.9 & 6.7   \\		 
neutral 		& 0 & 2290 & 2240 & 4.58 & 4.48 & 4.75 & 1.025 & (5.2,4.1) & 8.6 & 9.2 \\	 	 
unstable		& -40 & 1970 & 1910 & 6.74 & 6.69 & 7.75  & 1.037 & (30,15) & 11.3 & 12.7  \\ \hline
	\end{tabular}
	\caption{Simulation parameters and selected output variables. The output bulk Reynolds numbers $Re_b^I$ and $Re_b^{III}$ and Nusselt numbers  $Nu_b^I$ and $Nu_b^{III}$ are volume-averaged and time-averaged over 50 friction time units before the end of stages I and III, respectively. $q^s$ is the conductive heat flux through the ice imposed as an initial condition at the beginning of stage II. $\xi^{III}$, $\xi_-$ and $\xi_+$ are the mean interface position, the maximum amplitude of the keels and the maximum amplitude of the channels averaged over 50 friction time units at the end of stage III. $C_D^I$ and $C_D^{III}$ are the drag coefficients averaged over 50 friction time units at the end of stage I and stage III, respectively. Note that {$Re_*=150$}, $Pr=1$ and $St=1$ in all simulations. Also, $Ri_*=-40$ corresponds to $Ra=4.5\times 10^5 $. }	
	\label{table1}
\end{table}

We investigate the effect of background density stratification on the generation of topography at the fluid-solid interface by considering three distinct values of $Ri_*$, i.e., $Ri_*=40$, $Ri_*=0$ and $Ri_*=-40$, for which the stratification is stable, neutral and unstable, respectively. {For simplicity and computational expediency, all other parameters (except $T_t$) are fixed such that the flow is (moderately) turbulent and phase changes are relatively rapid, i.e., we set $Re_*=150$ ($Re=11250$), $Pr=1$ and $St=1$. For each $Ri_*$, we set $T_t$ such that the initial heat flux in the ice, $-T_t/2$, is almost equal to the heat flux in the fluid when there is no melting. As a result, the fluid-solid interface position does not move significantly in time (at least initially) and we can maximize numerical resolution around the interface with a fixed grid.} For reference, the Rayleigh number for the unstable stratification case ($Ri_*=-40$) is $Ra=4.5\times 10^5$, which is above the instability onset for Rayleigh-B\'enard convection rolls in the streamwise direction ($Ra\approx 1101$) of thermally-stratified plane Poiseuille flow \cite[][]{Chandrasekhar1961,Gage1968}. {Note that changing the sign of $Ri_*$ can be obtained by changing the sign of the thermal expansion coefficient $\alpha$, which indeed can be either positive or negative depending on the fluid state. We discuss the geophysical relevance of our choice of parameters, including $\alpha$, in \S\ref{sec:geophysics}.}

We solve equations \eqref{03} using the open-source pseudo-spectral direct numerical simulation (DNS) code DEDALUS \cite[][]{Burns2020}. We use 256 Fourier modes in the $x$ and $y$ directions and a compound Chebyshev basis with a total of 288 modes in the $z$ direction  unless stated otherwise. {The use of a compound Chebyshev basis allows to have a stretched grid in the vertical direction with refined (resp. coarse) resolution near the mean fluid-solid interface (resp. in the fluid bulk). Here, the Chebyshev collocation grid has a resolution equal to approximately 1/4th of a wall unit $\Delta z^+=1/Re_*\approx 0.0066$ at and around the fluid-solid interface and equal to about 1 wall unit in the fluid bulk, whereas the Fourier collocation grids have a uniform resolution of roughly 7 and 3.5 wall units in $x$ and $y$, which is within the recommended resolution for channel flow simulations (see, e.g., \cite{Moin1998}, and appendices \ref{sec:appA},\ref{sec:appC} for more details).} We use a two-step implicit-explicit Runge-Kutta scheme for time integration. The CFL condition is typically set to 0.2 in the transient initial stage and 0.4 later on. At statistical steady state, the time step is typically $10^3-10^4$ times smaller than the friction time scale $1/(Re_*Pr)$, which is equal (in terms of dimensional variables) to $H$ divided by the steady-state friction velocity. We run each simulation for about 4 diffusive time scales, or 600 friction time scales, which takes roughly 2 million time steps, such that the total cost of the simulations is on the order of 1 million CPU hours. Figure \ref{fig1}(a) shows a snapshot of the temperature field in the fluid (red colormap) and the solid (blue colormap), as well as the velocity vectors (arrows) at select locations for $Ri_*=0$. Figure \ref{fig1}(c) shows the variations of the phase field along the thick solid line drawn in figure \ref{fig1}(b). The transition from $\phi=1$ in the fluid to $\phi=0$ in the solid occurs over a very thin diffuse interface of thickness $ \approx 0.007 H$. Simulation parameters and output variables are provided in table \ref{table1}.

\subsection{Variables of interest}\label{sub:vars}
We define the friction velocity, the bulk velocity and the Nusselt number as{
\ba{}\label{definitions}
u_* = \langle \phi \rangle\sqrt{-\overline{\lp\tau_d+\tau_{\nu}+\tau_w\rp}}|_{z=1.5}, \quad u_b = \f{\langle u \rangle}{\langle \phi \rangle}, \quad Nu = \langle q \rangle, 
\ea
}respectively (cf. details in appendix \ref{sec:appB}), where overbar denotes the horizontal average, and $\langle \cdot \rangle\equiv \int_{\mathcal{V}}d\mathcal{V}/\mathcal{V}$ denotes the volume average, i.e., such that $\langle \phi \rangle$ is the mean fluid fraction {and $u_b$ is the bulk velocity of the fluid phase}. In equation \eqref{definitions}, $\tau_d$, $\tau_{\nu}$ and $\tau_w$ are the linear damping, viscous and Reynolds shear stresses, and $q=wT-\p_zT$ is the heat flux. At statistical steady state, the full shear stress $\overline{\tau}=\overline{\lp\tau_d+\tau_{\nu}+\tau_w\rp}$ is approximately a linear function of $z$ and $q$ is approximately depth invariant, in agreement with channel flow simulations of a pure fluid (cf. appendix \ref{sec:appB} for details on stresses and depth-independent variables using the phase-field method). {Since $\overline{\tau}$ is a linear function of $z$, $u_*$ can be estimated from the full shear stress as $\sqrt{-\overline{\tau}}$ at any depth as long as it is properly rescaled by the height at which it is estimated. Here, we use the shear stress at the top boundary $z=1.5$ in equation \eqref{definitions} for convenience but with pre-multiplying coefficient $\langle \phi \rangle \leq 1$, such that $u_*$ is truly the friction velocity at the mean interface position (cf. equation \eqref{definitions}).} We denote by $\xi$ the fluid-solid interface position, where
\ba{}
\xi(x,y) = \int_0^{l_z} \phi dz,
\ea 
such that $\overline{\xi}=\langle \phi \rangle$ (note that one could alternatively define $\xi$ such that it satisfies $\phi(z=\xi)=0.5$ or $T(z=\xi)=0$) and we denote the melt rate by $\dot{m}=\p_t\xi$. {The drag coefficient of the fluid-solid boundary is defined as the ratio of the dimensionless wall shear stress $u_*^2$ divided by the dynamic pressure $u_b^2/2$, i.e., 
\ba{}
C_D=2\lp\f{u_*}{u_b}\rp^2
\ea
\cite[][]{Garcia-Villalba2011,Pirozzoli2017}}. The temporal fluctuations of the variables of interest will be mainly reported in terms of the friction time $t_*=Re_*Prt$. Occasionally, we will show vertical profiles of variables in terms of the distance from the interface, which we denote by $\chi(t,x,y)=\xi(t,x,y)-z$.

\section{Results}\label{sec:results}

The key findings of our work are that {(i) streamwise topographical features emerge from uneven melting and freezing at a phase boundary when the flow is driven by a pressure gradient, and that (ii) the type of density stratification affects the characteristic amplitude and spanwise wavelength of the streamwise patterns}. Thus, after a discussion of the evolution of global flow variables in \S\ref{sub:stages}, we directly present the results of the topographical features generated at the fluid-solid boundary in \S\ref{sub:channels}. We then investigate the interplay between the turbulent flow, the topography and phase changes in \S\ref{sub:twoway}-\ref{sub:reverse}, and finally discuss the evolution of the mean interface position and the statistics of {melting and freezing} in \S\ref{sub:melting}.

\subsection{Simulation stages and global flow variables}\label{sub:stages}
\begin{figure}
\centering
\includegraphics[width=.95\textwidth]{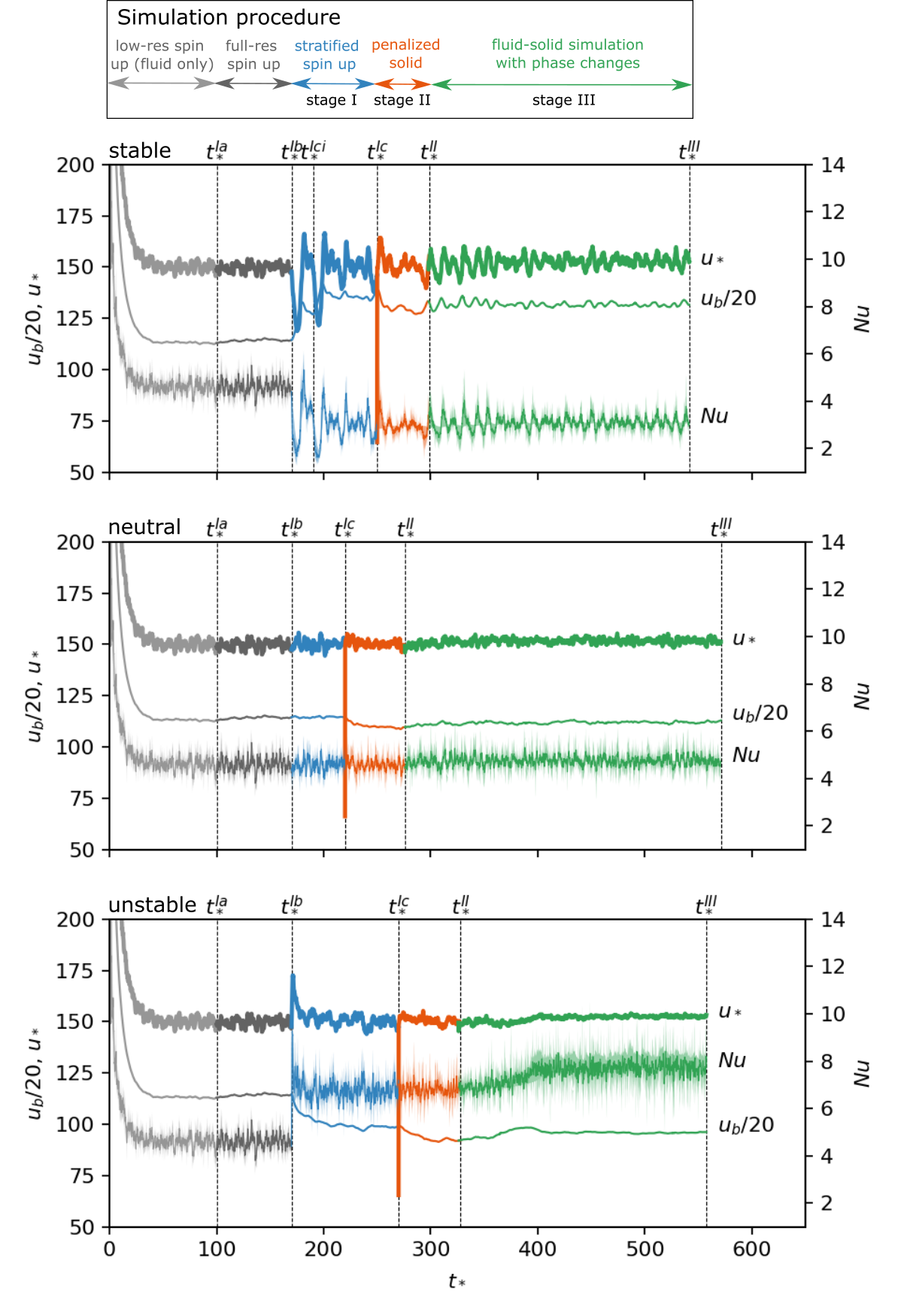}
\put(-360,468){\large{(a)}}
\put(-360,314){\large{(b)}}
\put(-360,160){\large{(c)}}
\vspace{-0.1in}\caption{{Time history of the friction velocity $u_*$ and bulk velocity $u_b$ (left axis), and of the Nusselt number $Nu$ (right axis) for (a) stable ($Ri_*=40$), (b) neutral ($Ri_*=0$) and (c) unstable ($Ri_*=-40$) density stratification. We report the friction time $t_*=Re_*Prt$ on the horizontal axis and show $u_b/20$ instead of $u_b$ as $u_b/20$ and $u_*$ have the same order of magnitude. Each full simulation can be broken down into a series of stages of increasing complexity, which are shown by different colors and are separated by vertical dashed lines. The overall procedure is shown schematically in the top panel.} There is an intermediate stage $t_*\in[t_*^{Ib},t_*^{Ici}]$ in (a) during which $Ri_*=20$. See the text and appendix \ref{sec:appC} for more details.}
\label{fig2}
\end{figure}

We show in figure \ref{fig2} the friction velocity $u_*$, the bulk velocity $u_b$ and the Nusselt number $Nu$ for stable (top figure), neutral (middle figure), and unstable stratification (bottom figure). Each simulation is broken down into three main stages, which are highlighted by different colors in figure \ref{fig2} {(note that we do not discuss the results shown by gray colors, which correspond to the spin-up of the fluid phase without buoyancy effects). The first stage  of interest (stage I for $t_*^{Ib} \leq t \leq t_*^{Ic}$) is shown by blue colors and corresponds to the spin-up of the fluid phase with buoyancy effects turned on. Importantly, stage I neglects the solid phase, which is substituted with a simple isothermal no-slip boundary, for computational expediency. The second key stage (stage II for $t_*^{Ic} < t \leq t_*^{II}$) is shown by orange colors and corresponds to the part of the simulations that includes the solid phase with volume penalization turned on, but neglects melting or freezing, i.e., such that the solid always occupies $1\leq z \leq 1.5$ and the phase field is prescribed as $\phi=0.5 \lcb 1 - \tanh\lb 2(z-1)/\delta \rb \rcb$ , where $\delta$ is the thickness of the diffuse interface. The final third stage (stage III for $t>t_*^{II}$) is shown by green colors and highlight results obtained when all effects are considered, i.e., buoyancy is turned on, there is both the fluid and the solid and phase changes are enabled (cf. additional details on the simulation stages in appendix \ref{sec:appC}).} The temperature in the solid is initialized at the beginning of stage II as
\ba{}
T=-q^s (z-1), \quad 1\leq z\leq 1.5,
\ea
where $q^s$ is the initial conductive heat flux through the solid, by imposing the fixed-temperature condition $T=T_t=-q^s/2$ at the top of the solid. The difference between the heat flux in the fluid and the conductive heat flux in the solid in stage II controls whether the fluid-solid interface melts or freezes once phase changes are turned on in stage III. Here, we set $q^s$ to be slightly smaller than the heat flux in the fluid at the end of stage I, which we denote $Nu^I$, such that the solid melts slowly at the beginning of stage III in all three simulations (see further discussion in \S\ref{sub:melting}). The bulk Reynolds and Nusselt numbers at the end of stages I and III are defined as
\ba{}\label{ReNu}
Re_b^I = \int_{t_*^{Ic}-\Delta_*}^{t_*^{Ic}} \f{u_bdt_*}{Pr \Delta_*}, \; Re_b^{III} = \int_{t_*^{III}-\Delta_*}^{t_*^{III}} \f{u_b dt_*}{Pr\Delta_*}, \; Nu^I = \int_{t_*^{Ic}-\Delta_*}^{t_*^{Ic}} \f{Nu dt_*}{\Delta_*}, \; Nu^{III} = \int_{t_*^{III}-\Delta_*}^{t_*^{III}} \f{Nu dt_*}{\Delta_*},
\ea
with $\Delta_*=50$ and are reported with $q^s$ in table \ref{table1}. {Note that $Re_b\ll Re$ because the flow is turbulent and hence experiences enhanced friction at the wall compared to the same flow in the laminar regime.}

Buoyancy effects are turned off for $t_* \leq t_*^{Ib}$, such that the results of figure \ref{fig2} are exactly the same for all three simulations until $t_* = t_*^{Ib}$. Upon turning on buoyancy, i.e., for $t_* \geq t_*^{Ib}$ (blue colors), the Nusselt number and bulk velocity deviate from the neutral case (middle figure), but with opposite behaviors: $Nu$ decreases while $u_b$ increases with stabilizing buoyancy effects (top figure), and $Nu$ increases while $u_b$ decreases with destabilizing buoyancy effects (bottom figure). The friction velocity, on the other hand, remains close to $u_*/Pr\approx Re_*$ in all three cases. The effect of background stratification on bulk velocity and heat fluxes are well known from channel-flow studies {\cite[][]{Garcia-Villalba2011,Pirozzoli2017}}, and the important point is that the heat flux is the variable that changes the most with buoyancy effects. Here, $Nu^I=3.02$, 4.58 and 7.75 for $Ri_*=40$, 0 and -40 respectively (cf. table \ref{table1}). It is worth noting that while $Nu$ remains the same between stage I and stage II (in a time-average sense), $u_*$ and $u_b$ show some variations as a result of turning on volume penalization and adding a solid phase. The large dip of $u_*$ at $t_* \approx t_*^{Ic}$ is merely the result of a sudden deceleration of the mean flow close to the interface, due to the addition of linear damping, which is transient, as can be seen from the rapid return of $u_*$ to its statistically steady state value of $u_*\approx 150$. The drop of the bulk velocity is similarly due to the added linear damping. However, unlike the dip in $u_*$, the drop in $u_b$ persists at all times, implying that volume penalization results in anomalous drag on the mean flow. Here, the relative drop of bulk velocity is on the order of 5\% and the profiles of temperature and velocity close to the fluid-solid interface in stage II reproduce closely those obtained in stage I (see appendix \ref{sec:appA}). Therefore, we consider the discrepancy to be small enough not to warrant a computationally costly increase in resolution or further tuning of the phase-field parameters.

When melting is turned on, i.e., for $t_*>t_*^{II}$ (green colors), global flow variables show different behavior depending on $Ri_*$. For the stable case, $u_*$, $u_b$ and $Nu$ exhibit moderately-large fluctuations (as in previous stages), but do not exhibit any time-mean deviation (top figure). For the neutral case, we find a small increase in $u_*$, $u_b$ and $Nu$ (middle figure). For the unstable case, we find that {$u_*$ and $u_b$ increase slightly}, while $Nu$ increases substantially (bottom figure). The analysis presented in the next sections explains these behaviors. Eventually, all simulations reach a statistical steady state.

\begin{figure}
\centering
\includegraphics[width=0.7\textwidth]{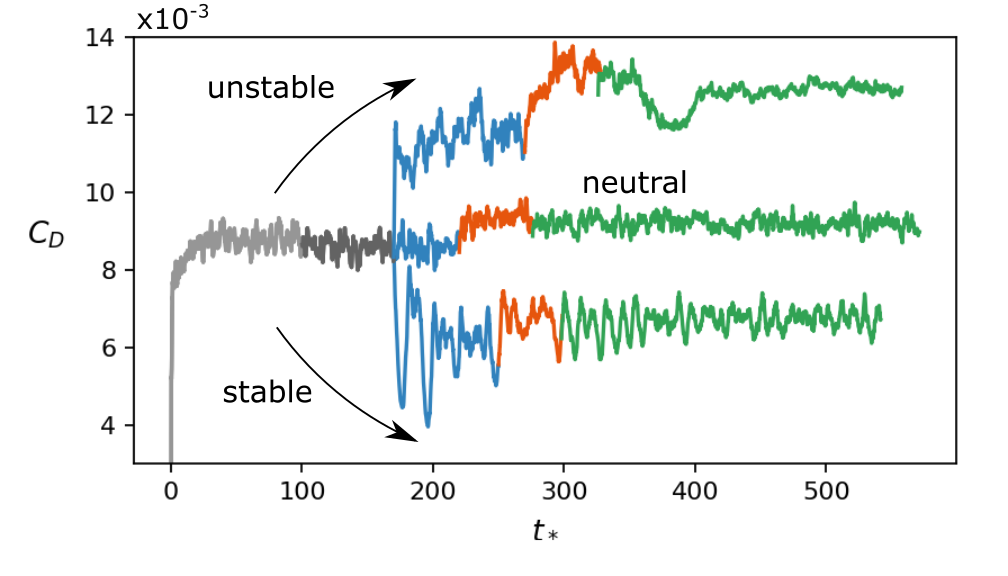}
\vspace{-0.1in}\caption{Drag coefficient $C_D$ as a function of time $t_*$ for unstable, neutral and stable stratification. The colors highlight different simulation stages as in figure \ref{fig2}.}
\label{fig11}
\end{figure}

We show in figure \ref{fig11} the temporal evolution of another global variable, namely, the drag coefficient, $C_D$, which is of significant interest in inferring melt rates from resolved variables in coarse models (using, for instance, the three-equation model; see, \cite{Holland1999}). The drag coefficient {is of order $10^{-2}$} and decreases (resp. increases) significantly at $t_*=t_*^{Ib}$, i.e., when the stratification becomes stable (resp. unstable). {The decrease (resp. increase) of $C_D$ results from an increase (resp. decrease) of the potential energy barrier in stirring the mean shear and bringing momentum upward with increasing stable (resp. unstable) stratification and is in agreement with previous studies \cite[][]{Garcia-Villalba2011,Pirozzoli2017}}. In stage II, $C_D$ increases because $u_b$ decreases moderately upon turning on volume penalization (cf. figure \ref{fig2}). In stage III, $C_D$ has similar values as in stages I and II (cf. reported values in table \ref{table1}), showing that it is not modified by the topographical features obtained in DNS, perhaps because they are aligned with the main flow direction (see section \ref{sub:channels}).

\subsection{Spontaneous generation of channels and keels}\label{sub:channels}

The mean interface position does not vary significantly in our simulations, due to our choice of initial and boundary conditions for the solid, but uneven melting by the turbulent flow still generates large-amplitude topography, which we discuss in this section. We denote variables averaged in the $x$ direction by a tilde ( $\widetilde{\;}$ ) and variables averaged in the $x$ direction minus the horizontal mean by a prime ( $'$ ), i.e., such that, e.g., $\xi'=\widetilde{\xi}-\overline{\xi}$ represents the spanwise variations of the streamwise-averaged topography around the horizontal mean.

\begin{figure}
\centering
\includegraphics[width=1\textwidth]{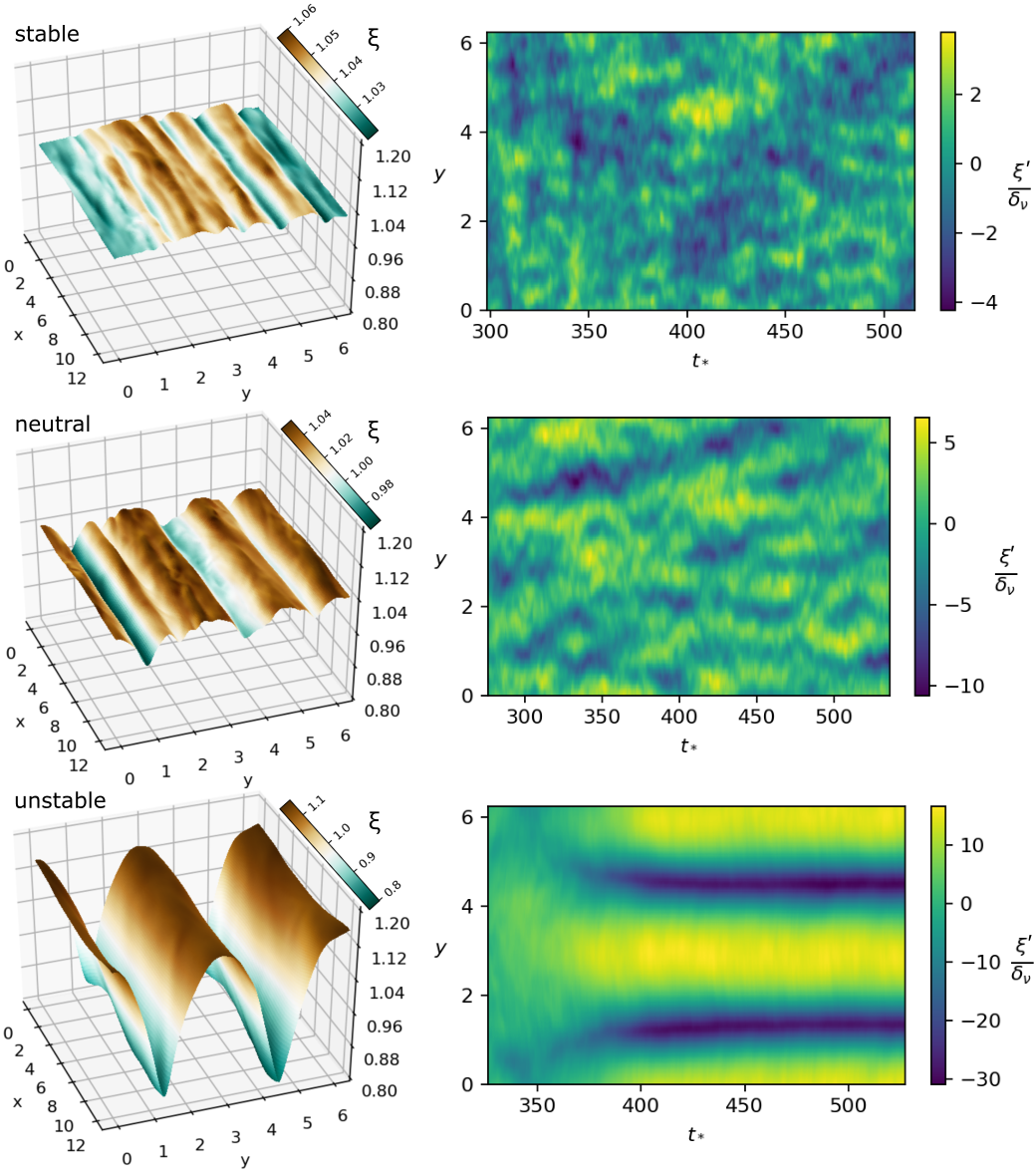}
\put(-395,425){\large{(a)}}\put(-235,425){\large{(d)}}
\put(-395,280){\large{(b)}}\put(-235,280){\large{(e)}}
\put(-395,135){\large{(c)}}\put(-235,135){\large{(f)}}
\vspace{-.in}\caption{(a-c) Interface topography between the fluid and the solid at the final time of the simulation. Channels in the ice are highlighted by brown colors while keels are highlighted by green colors. (d-f) Spanwise variations of interface topography $\xi'=\widetilde{\xi}-\overline{\xi}$ normalized by the viscous length scale $\delta_{\nu}=1/Re_*$ and as functions of $t_*$. The top, middle and bottom rows show the results for stable, neutral and unstable stratification, respectively.}
\label{fig3}
\end{figure}
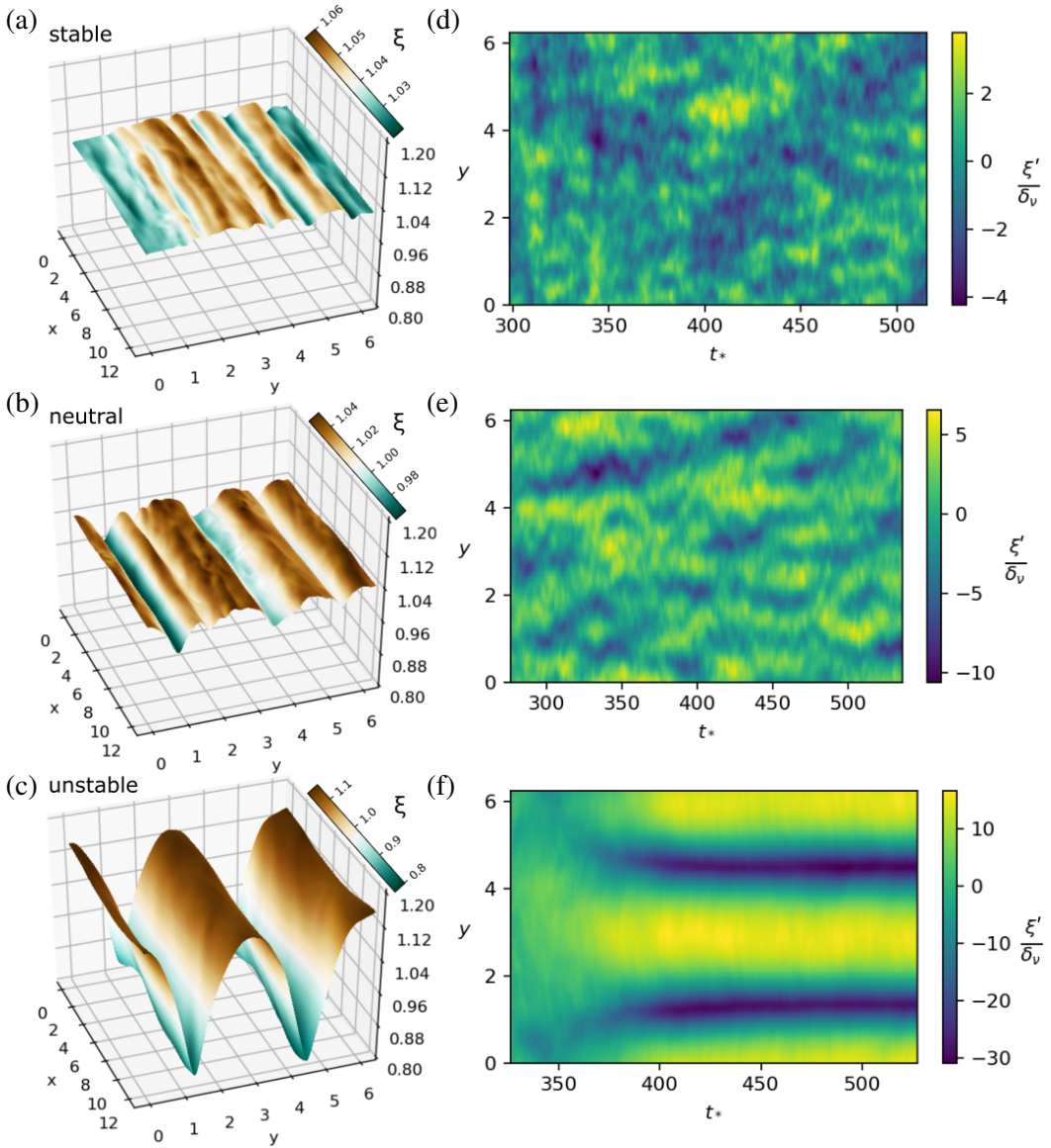

We show snapshots of the two-dimensional fluid-solid interface $\xi$ at the end of stage III in figures \ref{fig3}(a-c) for stable, neutral and unstable stratification, respectively. In all three cases, the topography is dominated by channels (troughs in the solid; brown colors), and keels (excursions of solid into the fluid; green colors), aligned with the streamwise direction. {We reach a statistical steady state relatively quickly in all cases after turning on phase changes, such that the patterns in figure \ref{fig3}(a-c) are representative of the interface topography throughout most of stage III (see Movies 1-3 in Supplementary Material). We show in figures \ref{fig3}(d-f) the H\"{o}vmoller diagrams of the channels and keels by plotting $\xi'$ in $(t_*,y)$ plane for all of stage III. It can be seen that the characteristic amplitudes of the channels and keels saturate almost immediately for stable and neutral stratification and well before the end of stage III for unstable stratification.} The steady-state amplitude of the biggest channels, $\xi_+$ (maximum of $\xi'$), and the steady-state amplitude of the deepest keels, $\xi_-$ (minus the minimum of $\xi'$), increase with decreasing $Ri_*$ (i.e., from figure (d) to (f)). The crest-to-trough amplitude is roughly 5, 10 and 45 times the viscous length scale $\delta_{\nu}=1/Re_*$ for stable, neutral  and unstable stratification, respectively (note that $\delta_{\nu}$ is roughly equal to the diffuse interface thickness; cf. appendix \ref{sec:appA}). Thus, the crest-to-trough amplitude is of the same order as the viscous sublayer thickness, which is approximately $5\delta_{\nu}$, for stable and neutral stratification, but extends beyond the buffer layer and into the log layer for the case of unstable stratification (figures \ref{fig3}(c),(f)).

Figures \ref{fig3}(d-f) show that the viability of channels and keels increases with decreasing stratification: channels and keels are short lived with stable stratification but long lived with unstable stratification. For stable stratification (figures \ref{fig3}(d)), the separation of scales between the topography lifetime (about 10 friction time units) and the diffusion time scale across the solid layer (about 100 friction time units), suggests that the interface evolution is purely driven by the flow dynamics. For neutral stratification, figure \ref{fig3}(e)) shows that channels and keels can drift, merge, split, decay and spontaneously appear over time scales of tens to hundreds of friction time units, highlighting a possible interplay between interface evolution and the fixed-temperature condition at the top solid boundary. For unstable stratification (figures \ref{fig3}(f)), the channels and keels become time invariant and their amplitudes saturate because of the top solid boundary condition, which plays a key role in the interface evolution as discussed in the next sections.

\subsection{Coupled dynamics of the fluid and solid phases}\label{sub:twoway}

The emergence of channels and keels can be the result of either (i) a passive response of the interface to uneven melting patterns driven by the turbulent flow, or (ii) a fully-coupled interplay between fluid turbulence, interface topography, and temperature in the solid. Here, we investigate the relevance of regimes (i) and (ii) for each of our simulations by looking at both the flow dynamics and the temperature field in the solid.

\begin{figure}
\centering
\includegraphics[width=0.95\textwidth]{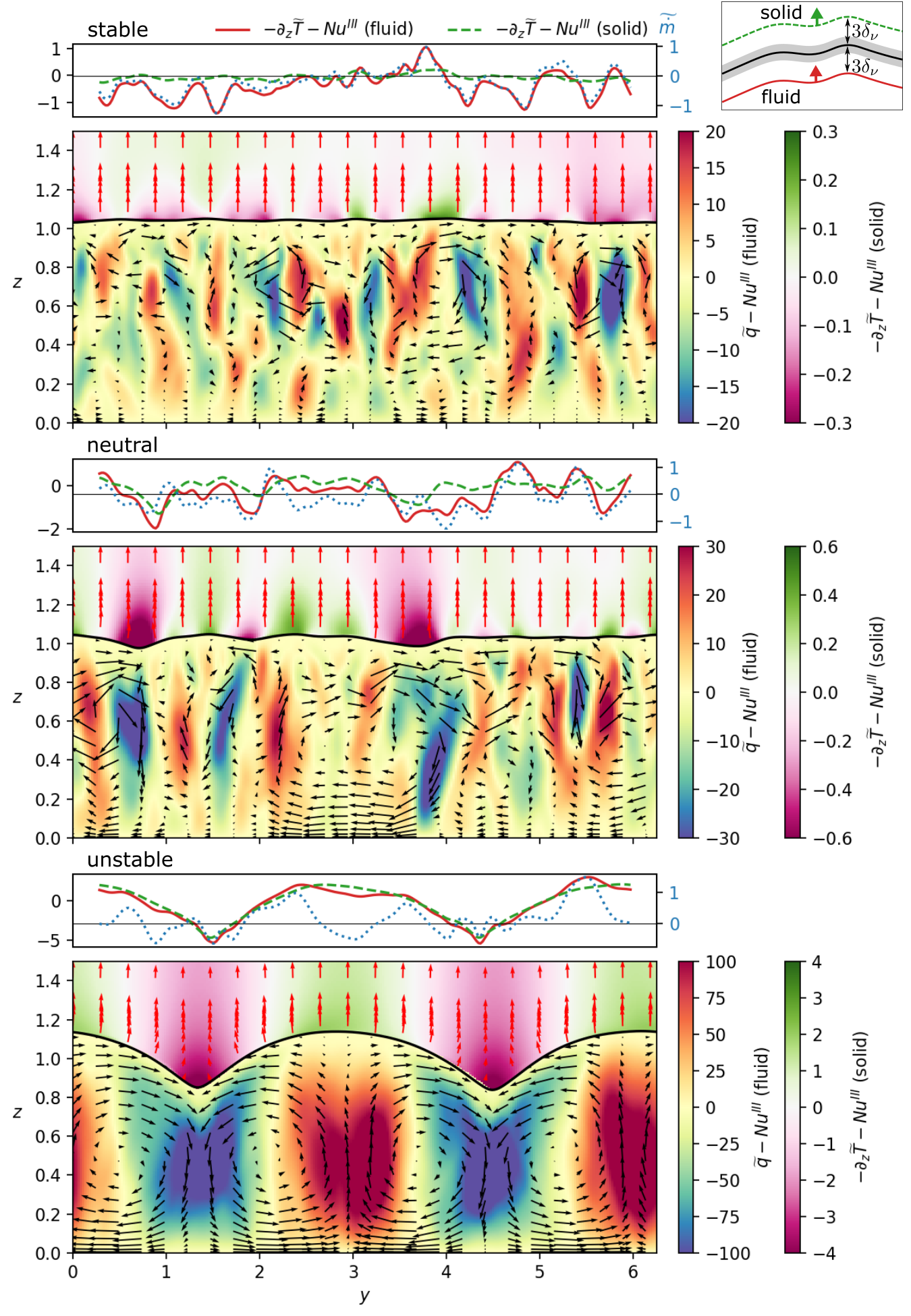}
\put(-365,508){\large{(a)}}
\put(-365,340){\large{(b)}}
\put(-365,172){\large{(c)}}
\vspace{-0.1in}\caption{{Panels of vertical heat fluxes for (a) stable, (b) neutral and (c) unstable stratification at time $t_*\approx 500$ representative of the statistical steady states. The lower figure of each panel shows the $x$-averaged vertical heat fluxes $-\p_z\widetilde{T}-Nu^{III}$ in the solid (pink-green colormap) and $\widetilde{q}-Nu^{III}$ in the fluid (spectral colormap), the $x$-averaged interface position $\widetilde{\xi}$ (black solid line), and the velocity vectors $(\widetilde{v},\widetilde{w})$ in the fluid (black arrows) and  the conductive heat flux $(-\p_y\widetilde{T},-\p_z\widetilde{T})$ in the solid (red arrows). The upper figure of each panel shows the vertical conductive heat flux $-\p_z\widetilde{T}-Nu^{III}$ at a distance $3\delta_{\nu}=0.02$ above (green dashed lines) and below (solid red lines) the interface, and the melt rate (blue dotted lines), as illustrated in the upper right panel where the gray shading highlights the thickness of the diffuse interface.}}
\label{fig4}
\end{figure}

{Figure \ref{fig4} shows the $x$-averaged vertical heat fluxes in both the fluid and solid for stable (top panel), neutral (middle panel) and unstable stratification (bottom panel) at a late time when statistical steady state is reached. In the lower figure of each panel, we show the heat map of the vertical heat flux $\widetilde{q}=\widetilde{wT}-\p_z\widetilde{T}$ in the fluid and of the conductive heat flux $-\p_z\widetilde{T}$ in the solid (note that we subtract $Nu^{III}$ in order to highlight fluctuations and that $-\p_z\widetilde{T}$ provides a more accurate measure than $\widetilde{q}$ for the heat flux in the solid; cf. appendix \ref{sec:appB}). In all three simulations, the spanwise fluctuations of the vertical heat flux are one order of magnitude (or more) larger in the fluid bulk where convection is active than in the solid where there is no movement. This suggests that the fluid flow has the capacity to induce transient melting or freezing hot spots and generate topography on short time scales, whereas temperature diffusion in the solid is moderate and primarily passive since it acts on long diffusive time scales of order 100 friction time units. The upper figure of each panel shows the melt rate $\widetilde{\dot{m}}$ as a function of $y$ (dotted blue line), as well as the vertical conductive heat fluxes in the fluid (solid red line) and solid (dashed green line) right below and above the interface, as illustrated in the top right panel. As expected, the melt rate is positive when the heat flux coming from the fluid exceeds the heat flux going through the solid, i.e., when the red solid line is above the green dashed line. The spanwise fluctuations of $-\p_z\widetilde{T}$ are moderately (resp. slightly) larger in the fluid than in the solid for stable (resp. neutral) stratifications. This suggests that the fluid dynamics controls the topography for stable and neutral stratification. In fact, we can observe that most local maxima of $-\p_z\widetilde{T}$ in the fluid near the boundary shown by the red solid lines in the line plots of figure \ref{fig4} correspond to large upwelling events in the fluid bulk (see, e.g., $y\approx 4$ for the top panel and $y\approx 0$ for the middle panel). However, heat fluxes in the solid still play a key role since there would be no freezing for a solid with a uniform temperature field. } Heat flux fluctuations increase in the solid phase as well as in the fluid phase with decreasing $Ri_*$ (i.e. from top to bottom row): enhancement of heat flux fluctuations are the result of buoyancy effects in the fluid and of larger interface deformation in the solid. All heat maps show that $-\p_z\widetilde{T}-Nu^{III}$ in the solid is generally positive (green) over a channel and negative (pink) over a keel. In the case of stable and neutral stratification, $-\p_z\widetilde{T}-Nu^{III}$ decays from large fluctuation values near the interface to almost 0 near the top boundary, suggesting no significant influence of the top boundary on the temperature field in the solid (i.e., increasing the ice thickness while adjusting the top temperature to conserve the heat flux would not change the results). On the other hand, in the case of unstable stratification (figure \ref{fig4}(c)), {$-\p_z\widetilde{T}$ fluctuations} remain large near the top boundary, suggesting that there is a backreaction from the fixed-temperature top-boundary condition, $T=-q^s/2$ at $z=1.5$, on the interface evolution. The backreaction from the top boundary for unstable stratification is obtained because the position of the channels and keels becomes rapidly stationary (contrary to the stable and neutral cases), such that the temperature field in the solid has time to adjust diffusively and balance the growth of the channels and keels (cf. appendix \ref{sec:appD} for more details on the temperature field in the solid). {The steadiness of the fluid dynamics and interface topography for unstable stratification can be seen to result in overlapping heat fluxes in the top figure of the bottom panel.} 

The emergence of streamwise channels and keels is consistent with the well-documented presence of near-wall streamwise streaks and vortices in stratified shear flows \cite[][]{Pirozzoli2017,Zonta2018}, but their amplitude clearly varies with $Ri_*$. In the case of stable stratification (figure \ref{fig4}(a)), buoyancy effects inhibit the generation of large topographical features, such that channels and keels have small amplitudes and do not feed back onto the flow (which is further discussed in \S\ref{sub:reverse}). In the case of neutral stratification, buoyancy is turned off, such that the solid boundary deforms more and can affect the flow dynamics. Figure \ref{fig4}(b) shows that the heat flux in the fluid close to the boundary is usually larger where there are channels (e.g. $y\approx 1.5,2.5$) than where there are keels (e.g. $y=0.9$), suggesting a topographic influence on the flow. For the case of unstable stratification (figure \ref{fig4}(c)), two streamwise rolls aligned with the direction of the flow and filling the entire depth dominate the fluid dynamics. These flow features are reminiscent of Rayleigh-B\'enard convection rolls as observed in channel flow simulations with unstable stratification \cite[][]{Pirozzoli2017}, which here appear locked within the interface deformation pattern (figure \ref{fig4}(c)). 

\begin{figure}
\includegraphics[width=1\textwidth]{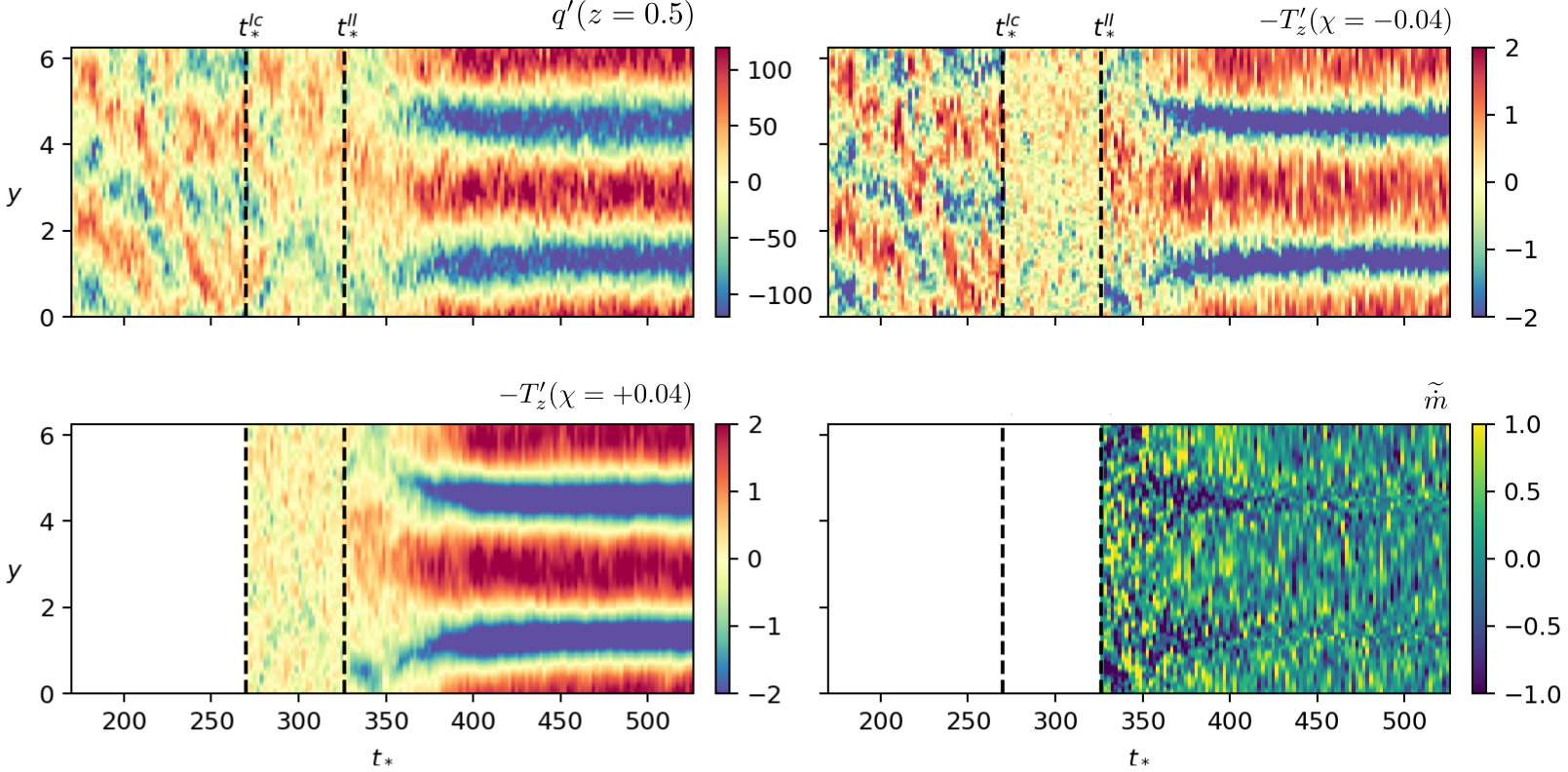}
\put(-380,183){\large{(a)}}
\put(-195,183){\large{(b)}}
\put(-380,90){\large{(c)}}
\put(-195,90){\large{(d)}}
\vspace{-0.05in}\caption{(a) Heat flux anomaly $q'=(wT)'-T_z'$ in the fluid at $z=0.5$ and {conductive heat flux anomaly $-T_z'$ (b) in the fluid at $\chi=-0.04$ (6 wall units) below the interface and in the solid at $\chi=+0.04$ above the interface} as functions of $(t_*,y)$ for the simulation with unstable stratification. The two vertical dashed lines highlight the times at which we turn on volume penalization and melting. {(d) shows the melt rate $\widetilde{\dot{m}}$ averaged in the streamwise direction.}}
\label{fig5}
\end{figure}

\begin{figure}
\centering
\includegraphics[width=0.5\textwidth]{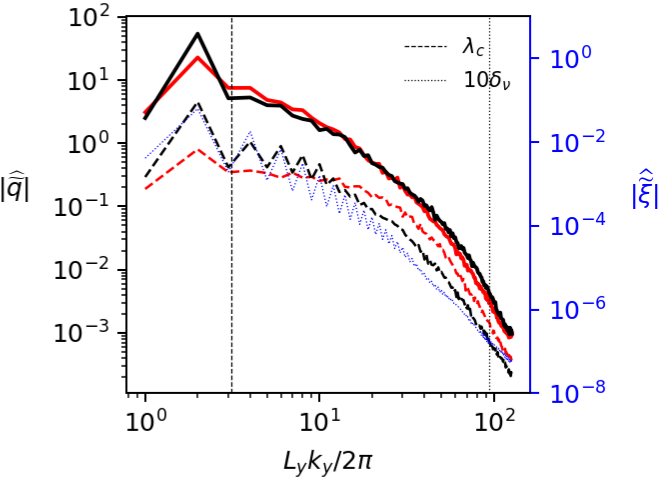}
\vspace{-0.05in}\caption{Spanwise spectrum of the heat flux $\widetilde{q}$ in the fluid at $z=0.5$ (solid lines) and at $\chi=-0.04$ below the interface (dashed lines). Black and red colors highlight the spectra obtained in stage I and III, respectively (with 100 friction time units averaging). {The blue dotted line shows the spectrum of $\widetilde{\xi}$ averaged over the last 100 friction time units of stage III.}}
\label{fig5b}
\end{figure}

The interplay between the solid boundary and the flow dynamics for the case of unstable stratification is further highlighted in figures \ref{fig5} and \ref{fig5b}. Figure \ref{fig5} shows the H\"{o}vmoller diagram of the heat flux in the middle of the fluid (figure \ref{fig5}(a)) {and at 6 wall units below (figure \ref{fig5}(b)) and above the interface (figure \ref{fig5}(c)).} Two mid-depth streamwise rolls whose positions are locked are evident from $t_*\approx 370$ onward in figure \ref{fig5}(a), which is about the same time as when the two channels and keels become large in figure \ref{fig3}(f). Similar rolls can be inferred for $t_*<370$ but are weaker and meander. { The two strong rolls for $t_*\geq 370$ are locked with the topography (cf. figure \ref{fig3}(f)) and support large conductive heat fluxes with similar patterns right below the interface (cf. figure \ref{fig5}(b)), which further demonstrates that global modes control the interface dynamics for an unstable stratification. The conductive heat fluxes coming from the fluid are yet eventually balanced by the conductive heat fluxes through the solid (figure \ref{fig5}(c)), which adjust diffusively as the interface deforms, such that there is no net melting (figure \ref{fig5}(d)) beyond the initial transient of stage III. Note that the decrease of conductive heat flux below the interface between $t_*^{Ic}$ and $t_*^{II}$ in figure \ref{fig5}(b) is due to the heat flux imbalance at the beginning of stage II (cf. appendix \ref{sec:appA}) and has no incidence on the subsequent melting dynamics.} Figure \ref{fig5b} shows the spanwise spectrum of $\widetilde{q}$ (left axis) in the middle of the fluid (solid lines) and near the solid boundary (dashed lines) at the end of stages I (red colors) and III (black colors) for the unstable case. {In all cases, the spectrum peaks at wavenumber $L_yk_y/2\pi=2$, which is close to the critical wavenumber $L_yk_c/2\pi= 3.11$ of convection instability ($\lambda_c=2.016$),} suggesting that Rayleigh-B\'enard convection is already active in stage I. However, with melting turned on, the peak is significantly amplified, especially for the spectrum near the boundary (dashed lines), which also shows amplification of higher harmonics, consistent with the spectrum of the interface itself (dotted blue line; right axis). These result suggest that Rayleigh-B\'enard rolls are energized more than any other fluid features once melting is turned on, because they best couple with the interface topography evolution as a result of melting and freezing. Note that the spectra of the heat flux near the boundary and of the interface have a sawtooth-like pattern due to the non-sinusoidal shape of the interface and numerical confinement in the spanwise direction.

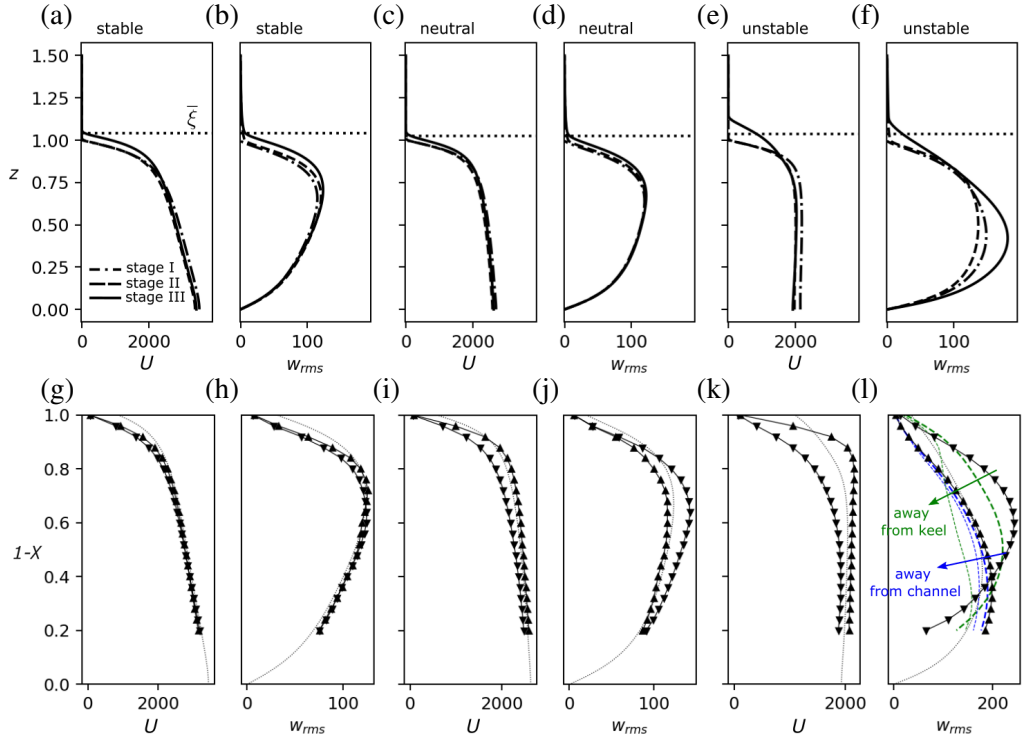
\begin{figure}
\centering
\includegraphics[width=1\textwidth]{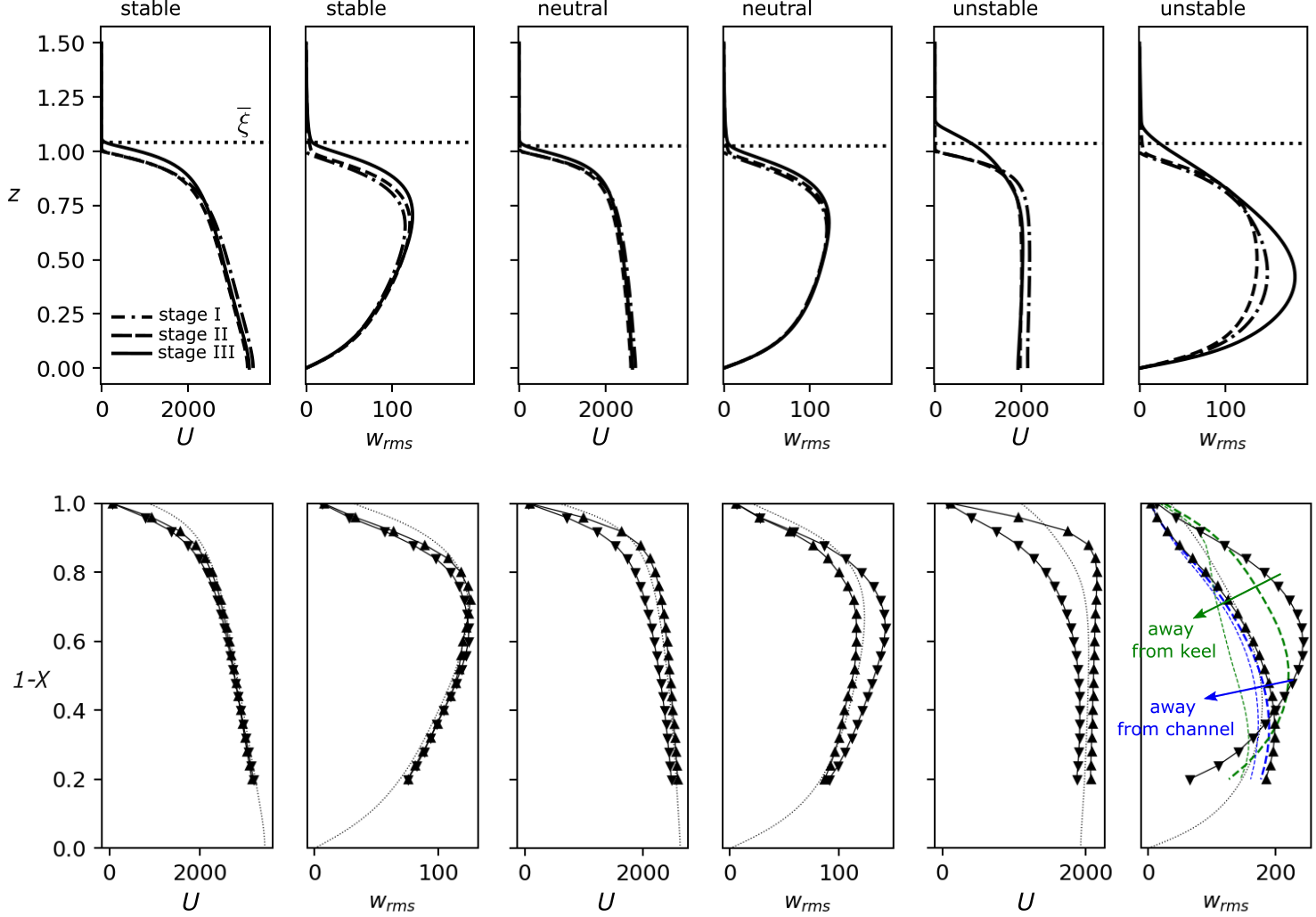}
\put(-370,267){\large{(a)}}
\put(-308,267){\large{(b)}}
\put(-245,267){\large{(c)}}
\put(-185,267){\large{(d)}}
\put(-123,267){\large{(e)}}
\put(-65,267){\large{(f)}}
\put(-370,127){\large{(g)}}
\put(-308,127){\large{(h)}}
\put(-245,127){\large{(i)}}
\put(-185,127){\large{(j)}}
\put(-123,127){\large{(k)}}
\put(-65,127){\large{(l)}}
\vspace{-0.0in}\caption{{Vertical profiles of the streamwise velocity $U$ and vertical velocity rms $w_{rms}$ averaged in $x$ and $t_*$ (over at least 50 friction time units) for stable (two leftmost columns), neutral (two middle columns) and unstable stratification (two rightmost columns). The top panels show the full horizontal average of $U$ and $w_{rms}$, i.e., when they are also averaged in $y$, and the mean interface position $\overline{\xi}$ at the end of stage III (dotted lines). The bottom panels show the full horizontal averages of $U$ and $w_{rms}$ as well (dotted lines), but also $U$ and $w_{rms}$ under the largest keel (downward triangles) and channel (upward triangles), i.e., such that they are not averaged in $y$, as functions of $1-\chi$, where $\chi$ is the distance from the interface. In (l), the green (resp. blue) thick and thin dashed lines show the vertical profiles shifted away from the largest keel (resp. channel) by 0.25 and 0.5 units in the $+y$ direction.}}
\label{fig6}
\end{figure}

We next show in figure \ref{fig6}(a-f) (top row) the vertical profiles of the mean streamwise velocity $U$ and of the root-mean-square (rms) vertical velocity, $w_{rms}$, where here mean and rms are defined using a horizontal and temporal average. The results are shown for all three stages and for stable (two left-most columns), neutral (two middle columns) and unstable stratification (two right-most columns). The vertical profiles are never symmetric with respect to the half fluid depth position, which is $z=0.5$ in stages I and II. This is because our velocity boundary conditions across the fluid layer are different, i.e., no-slip on the top boundary (which can be the fluid-solid interface) and free-slip on the bottom. For stable and neutral conditions, there is a strong overlap of all curves, suggesting that the topography doesn't influence the mean profiles, while for unstable stratification, there is a small deviation of the stage III profiles (solid lines). {Note that a log-log plot of the mean velocity and temperatures profiles near $z=1$, which we show in figure \ref{figapp2} in appendix \ref{sec:appA}, clearly shows that the flows follow the law of the wall on the top no-slip fluid boundary in stages I and II.}

We investigate in figure \ref{fig6}(g-l) whether the mean profiles vary in the spanwise direction in such a way that they correlate with the $x$-averaged interface topography. To do so we consider the mean profiles under the largest keel (downward triangles), i.e., at $y=y_k$ where $y_k$ is the minimum in $y$ of $\widetilde{\xi}$ at each time step, separately from the mean profiles under the largest channel (upward triangle), i.e., at $y=y_c$ where $y_c$ is the minimum in $y$ of $\widetilde{\xi}$ at each time step, where mean now denotes streamwise and temporal averaging. For stable stratification, there is no difference between the profiles under keels and channels. However, for neutral and unstable stratification, the two profiles depart in such a way that the streamwise velocity is larger under channels than under keels, and the vertical velocity rms (defined using a temporal and $(x,y)$ average for each $\chi$) is larger under keels than under channels. These results indicate a noticeable influence of the topography on the flow. For the case of neutral stratification, the separation of the profiles is maximum for $\chi<0.3$ and then vanishes, suggesting a local influence of the topography on the flow dynamics, while for unstable stratification the effect of the topography is felt throughout the entire depth due to coupling with the Rayleigh-B\'enard rolls. It may be noted that the profiles of $w_{rms}$ under the largest keels and channels are larger than the plane-average profile shown by the dotted line in figure \ref{fig6}(l). This is expected because Rayleigh-B\'enard convection promotes both localized intense upwellings and intense downwellings under channels and keels. In fact, away from the main channel and keel the profiles decrease rapidly, as can be seen from the blue and green lines.

\begin{figure}
\centering
\includegraphics[width=1\textwidth]{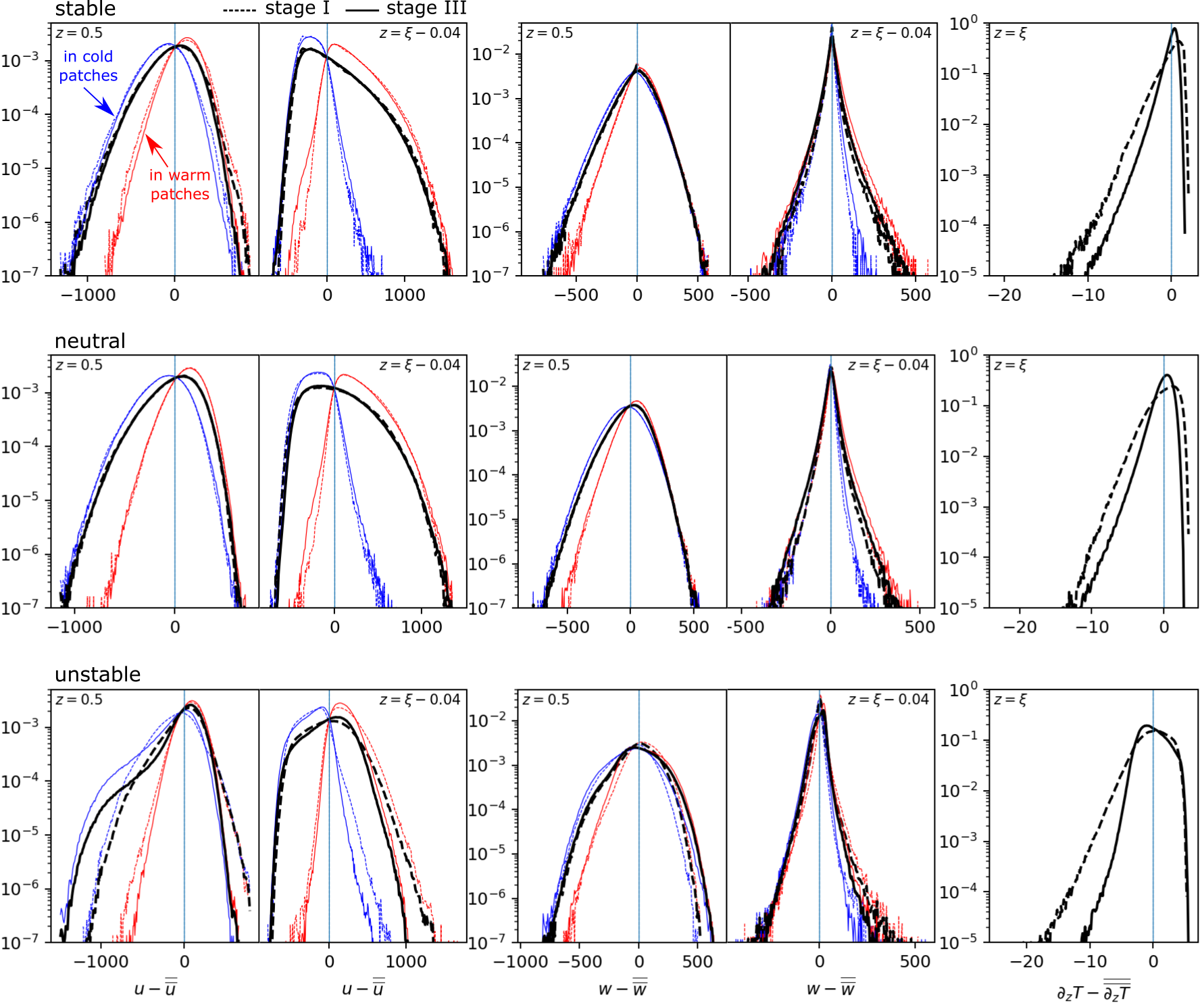}
\put(-380,320){\large{(a)}}\put(-230,320){\large{(b)}}\put(-80,320){\large{(c)}}
\put(-380,213){\large{(d)}}\put(-230,213){\large{(e)}}\put(-80,213){\large{(f)}}
\put(-380,106){\large{(g)}}\put(-230,106){\large{(h)}}\put(-80,106){\large{(i)}}
\vspace{-0.0in}\caption{{Probability density functions (pdfs) of the streamwise velocity fluctuations $u-\overline{\overline{u}}$ (leftmost panels), vertical velocity fluctuations $w-\overline{\overline{w}}$ (middle panels) and temperature gradient fluctuations $\p_zT-\overline{\overline{\p_zT}}$ (rightmost panels), for stable (top row), neutral (middle row) and unstable density stratification (bottom row), with double overbar denoting the horizontal and temporal average. The pdfs of $u-\overline{\overline{u}}$ and $w-\overline{\overline{w}}$ are shown at $z=0.5$ and $z=\xi-0.04$ (6 wall units below the interface), while the pdfs of $\p_zT-\overline{\overline{\p_zT}}$ are shown at the interface $z=\xi$. We use dashed lines and solid lines to highlight the results obtained in stage I and stage III, respectively. Blue and red colors show the pdfs of $u-\overline{\overline{u}}$ and $w-\overline{\overline{w}}$ conditioned on negative and positive local temperature fluctuations, in order to separate the pdfs of warm fluid patches from those of cold fluid patches.}}
\label{fig7}
\end{figure}
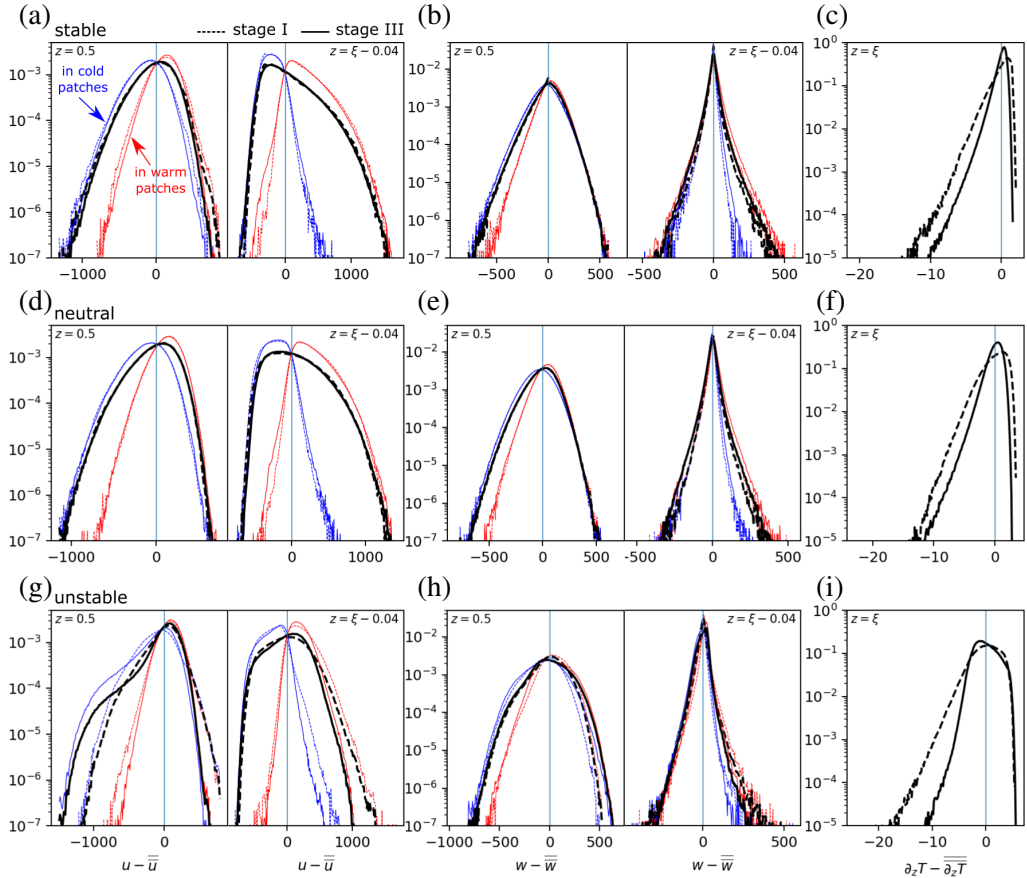

In order to gain further insight into the statistics of the flow interacting with the melting boundary, we show in figure \ref{fig7} the probability density functions (pdfs) of the streamwise velocity (left panels), the vertical velocity (middle panels) and the temperature gradient (right panels), for stable (top row), neutral (middle row) and unstable stratification (bottom row). For the velocities, the pdfs are shown both in the middle of the fluid, at $z=0.5$, and near the boundary, at $z=\xi-0.04$ (i.e., 6 wall units into the fluid). We find little difference between stage I (dashed lines) and III (solid lines) for the streamwise and vertical velocities, suggesting limited influence of the topography on the overall flow morphology, although the streamwise velocity in figure \ref{fig7}(g) (left panel) has a negative tail with higher probability density in stage III than in stage I. The temperature gradient at the {fluid-solid} interface (figures \ref{fig7}(c,f,i)) does vary noticeably between stage I and III. However, this difference is due to the phase-field method rather than to a fundamental change in flow morphology since the pdfs in stage III and II (not shown) show significant overlap. 

{While phase changes and the emergence of topographical features have little effect on the pdfs, most pdfs display flow-driven left-right asymmetries, which are worth highlighting. Most importantly, the pdf of the temperature gradient at the fluid-solid interface has a rapidly-decaying positive tail and a slowly-decaying negative tail. This asymmetry is obtained in all stages, hence is a feature of the flow rather than a consequence of topography generation, and suggests that the phase change dynamics should be itself asymmetric (which we show in section \ref{sub:melting})}. The pdfs of the streamwise velocity near the boundary are also asymmetric, featuring a slowly-decaying positive tail and a rapidly-decaying negative tail. We have further separated the pdfs of the velocities in figure \ref{fig7} based on the sign of the local temperature anomaly (compared to the plane and temporal mean). Blue curves denote pdfs obtained for negative temperature anomaly, i.e., representative of fluid patches influenced by the cold top boundary, while red curves denote pdfs obtained for positive temperature anomaly, i.e., representative of fluid patches influenced by the warm bottom boundary. The cold-temperature pdfs are shifted to the left of the warm-temperature pdfs for the streamwise velocity (left column), which suggests that negative streamwise velocity is more often associated with cold fluid coming from the top boundary. Also, for the vertical velocity at $z=\xi-0.04$ (right panels of the middle column), the positive tail of the warm-temperature pdfs (red) is larger than the negative tail of the cold-temperature pdfs (blue), suggesting more extreme warm upwelling events than cold downwelling events just outside of the viscous sublayer. These two results demonstrate that the near-wall flow dynamics have multiple asymmetries, which may be related to the asymmetry in the temperature gradient at the boundary.

\subsection{Reversing the stratification}\label{sub:reverse}

In the case of stable stratification, the cool melt fluid (at freezing temperature) is more buoyant than the warmer surrounding fluid, such that it rises and accumulates in the middle of the channels. Thus, channels and keels are limited to small amplitudes in figure \ref{fig4}(a) because of a negative feedback between channel generation and the melt dynamics, which halts channel growth. In the case of unstable stratification the opposite is true, i.e., the  cool melt fluid is denser than the surrounding fluid, hence is evacuated from channels. This yields a positive feedback between channel generation and melt dynamics, which results in the large-scale topographical features seen in figure \ref{fig4}(c) (which saturate over time due to thermal adjustment in the solid). 

\begin{figure}
\centering
\includegraphics[width=0.9\textwidth]{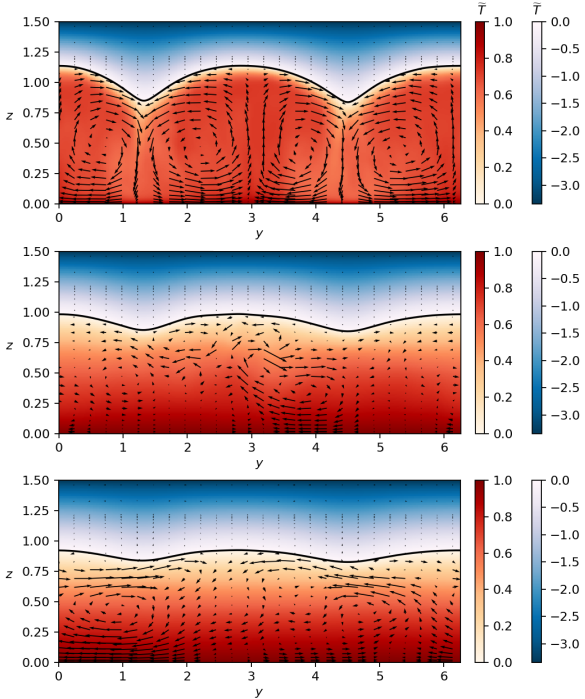}
\put(-350,400){\large{(a)}}
\put(-350,265){\large{(b)}}
\put(-350,130){\large{(c)}}
\vspace{-0.in}\caption{Temperature field averaged in $x$ direction in the fluid (red colormap) and in the solid (blue colormap) at times (a) $t_*=527$, (b) $t_*=539$, (c) $t_*=547$ for the restratifying simulation. The black arrows show the velocity field $(\widetilde{v},\widetilde{w})$. The thick solid line shows the $x$-averaged interface position $\widetilde{\xi}$.}
\label{fig8}
\end{figure}

The hypothesis of the cool melt fluid pooling in channels and inhibiting their growth for stable stratification is difficult to verify with the results discussed previously because of the small interface deformation obtained for $Ri_*=40$. Therefore, we have run a fourth simulation starting from the final time of the simulation with unstable stratification (and large interface deformation), but with an increasing Richardson number such that the fluid becomes stably stratified and interacts (transiently) with the initially-large topographical features. We impose the stable stratification through several intermediate steps so that the flow doesn't relax to a laminar state. Specifically, we use
\ba{}
Ri_*(t_*) = -40 \lb 1 - f(t_*,5) \rb + 20 \lb 2 + f(t_*,12) + f(t_*,19) \rb, 
\ea
with $f(t_*,\tau_*)=\tanh( t_*-t_*^{III} - \tau_*)$, such that $Ri_*$ starts from $\approx -40$ at $t_*=t_*^{III}$ and reaches $\approx 40$ for $t_*>t_*^{III}+19$. We show the results of this run in figure \ref{fig8}, where blue/red colors highlight $x$-averaged temperature values in the solid/fluid phase, while arrows denote $x$-averaged velocity vectors $(\widetilde{v},\widetilde{w})$ in the $(y,z)$ plane. Figure \ref{fig8}(a) shows the results at time $t_*=t_*^{III}+1$ ($t_*^{III}=526$), i.e., when $Ri_*\approx -40$. The stratification is unstable such that the flow features strong upwelling of warm fluid below the channels and strong downwelling of cool fluid along and under the keels, i.e., akin to Rayleigh-B\'enard rolls locked into the deformed interface pattern. A pair of counter-rotating streamwise rolls is clearly visible below each one of the two channels. {These rolls persist until $Ri_*\approx 0$, which is in agreement with recent simulations of mixed convection that have shown that streamwise rolls extending throughout the full depth of a channel (without phase changes) are obtained for a wide range of negative Richardson numbers \cite[][]{Pirozzoli2017}.} Figures \ref{fig8}(b-c) show the results at times $t_*=t_*^{III}+13$ and $t_*=t_*^{III}+21$, i.e., when $Ri_*\approx 0.7$ and $Ri_*\approx 39$, respectively. At these times, the stratification is stable and the cool melt fluid produced at the keels converges toward the channels' centreline. {The Rayleigh-B\'enard large-scale rolls have vanished and are replaced with weaker vortices of finite vertical extent, which are most vigorous close to the interface where they are driven by the (positive) buoyancy anomaly of the melt fluid at the tip of the keels.} The heat flux through the fluid goes down and freezing occurs everywhere such that the solid front advances into the fluid. Importantly, freezing is faster in the channels because of the convergence of the buoyant cool melt fluid {and higher conductive heat fluxes in the solid (as the solid is thin above channels)}, which leads to rapid refreezing of the initial channels. {While $Ri_*>0$ increases, the properties of the boundary-attached vortices (e.g., vertical extent and intensity) are the result of a complex interplay between the amplitude of the interface topography and of the stratification strength. The increasing stratification drives an increasing positive buoyancy anomaly of the melt fluid but also increasingly damps global modes \cite[][]{Garcia-Villalba2011}, while the decreasing topography amplitude is expected to result in flattening and weakening vortices. Nevertheless, the topography ultimately disappears and the vortices weaken significantly.}

\subsection{Melting}\label{sub:melting}

\begin{figure}
\includegraphics[width=1\textwidth]{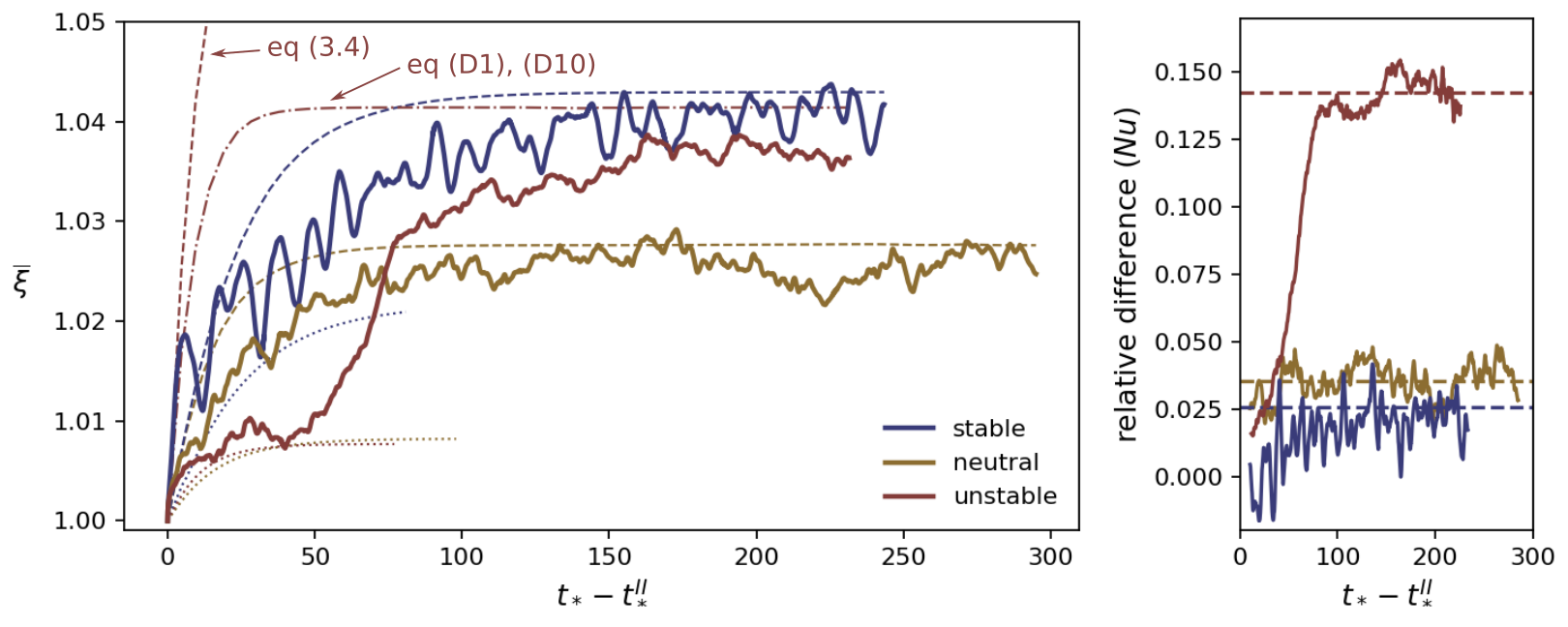}
\put(-370,155){\large{(a)}}\put(-100,155){\large{(b)}}
\vspace{-0.09in}\caption{(a) Time history of the mean interface position for stable, neutral and unstable stratification (thick solid lines). The dotted (resp. dashed) lines show the position of the interface predicted by the reduced model \eqref{toy} with $Nu^I$ (resp. $Nu^{III}$) as heat flux in the fluid. The dash-dot line shows the result of the higher-order model \eqref{D1}, \eqref{D10} for unstable stratification. (b) Time history of the heat flux relative to $Nu^I$, i.e., $(Nu-Nu^{I})/Nu^I$ (solid lines). We use a rolling mean over a $\Delta_*=20$ friction time units window in order to remove some of the most rapid large-amplitude oscillations. The horizontal dashed lines show $(Nu^{III}-Nu^{I})/Nu^I$.}
\label{fig9}
\end{figure}

In this section we discuss the evolution of the mean interface position $\overline{\xi}$ with time and the statistics of melting $\dot{m}=\p_t\xi$ at statistical steady state. We first show the evolution of $\overline{\xi}$ in DNS as a function of time in figure \ref{fig9}(a) for all three simulations (solid lines). {For a solid with spatially-uniform temperature equal to the melting temperature $T_m$, we would expect a faster increase of $\overline{\xi}$ with time for an unstable than for a stable stratification, since the heat flux in the fluid is larger when buoyancy forces are destabilizing instead of restoring. However, as indicated in section \S\ref{sec:model}, we have imposed a conductive heat flux in the solid ($q^s$) slightly less than the mean heat flux through the fluid ($Nu^I$) at the beginning of stage III, i.e., when we turn on melting, such that the leading-order melt rate is not controlled by the Nusselt number of the fluid-only simulations but by the difference $Nu^I-q^s$. This difference is 0.19, 0.1, 0.05, for stable, neutral and unstable stratification (cf. table \ref{table1}), such that the initial increase of $\overline{\xi}$ is faster for stable than for unstable stratification. Figure \ref{fig9} shows that $\overline{\xi}$ saturates over time. This happens because the conductive heat flux in the ice increases as $\overline{\xi}$ increases (in a plane-averaged sense), since the solid becomes thinner, such that it eventually balances the heat coming from the fluid.} 

Under the assumption of small interface deformations, it is possible to predict the evolution of the mean interface position over time using a reduced model. As a first approximation, we consider that the topography has no effect on the temperature in the solid, i.e., we assume that the heat flux through the solid is simply equal to the temperature difference between the interface and the top boundary divided by the mean solid thickness $h$ (cf. details in appendix \ref{sec:appD}, which hinges on an assumption of quasi steady state). Then, the evolution equation for $\overline{\xi}$ becomes
\ba{}\label{toy}
\f{d\overline{\xi}}{dt} = q^f - \f{h_0q^s}{h},
\ea
where $h_0=1/2$ is the initial ice thickness, $q^f$ is the heat flux in the fluid, and we recall $\overline{\xi}=\xi_0=1$ at $t=t_*^{II}$. For simplicity we take $q^f$ to be a constant diagnosed from the simulations. The results of equation \eqref{toy} for $q^f=Nu^I$, i.e., obtained when setting $q^f$ to the average heat flux before melting is turned on, are shown by the dotted lines in figure \ref{fig9}(a). The overlap between the reduced model and DNS results at early times is good for unstable stratification (as expected) but is poor for stable and neutral stratification. The disagreement with $q^f=Nu^I$ at early times arises because the temperature in the solid is slightly above 0 because of volume penalization, such that there is some artificially large melting at the beginning of stage III (cf. appendix \ref{sec:appA}). At later time, the DNS results and the model results shown by the dotted lines diverge because the heat flux $Nu$ increases rapidly once melting is turned on, as can be seen in figure \ref{fig9}(b). For unstable stratification, the agreement with $q^f=Nu^I$ is relatively good until $t_*\approx t_*^{II}+50$ (cf. red dotted line), i.e., right until the Rayleigh-B\'enard rolls are energized and a large-scale topography emerges (cf. \S\ref{sub:twoway}). 

In order to account for the increase in heat flux through the fluid enabled by melting and the generation of topography we show with dashed lines in figure \ref{fig9}(a) the result of equation \eqref{toy} with $q^f=Nu^{III}$, which is the heat flux at statistical steady state with melting turned on. For stable and neutral stratification, there is a good agreement between the model results and the mean interface position at late times. For unstable stratification, however, equation \eqref{toy} with $q^f=Nu^{III}$ overestimates the final value of $\overline{\xi}-1$ by a factor two (approximately), suggesting that topography plays a non-negligible role on the heat flux in the solid. We show in figure \ref{fig9}(a) a prediction of $\overline{\xi}$ for unstable stratification obtained using a more accurate higher-order model (red dash-dot line), which takes into account interface deformation (cf. appendix \ref{sec:appD}). The higher-order prediction overlaps well with the DNS results at late times, demonstrating that melting and the generation of topography changes the heat flux through both the fluid and the solid. {The topography makes the solid more efficient at evacuating heat because the anomalous (increased) heating obtained above the channels (i.e., where the solid is thin) exceeds in absolute value the anomalous (reduced) heating obtained above the keels (i.e, where the solid is thick), which is a nonlinear effect in the topography amplitude obtained for any topography with top-down symmetry (e.g., a sinusoid). The higher-order model takes into account this nonlinear effect in topography amplitude and predicts a steady-state solid layer thickness larger than that predicted by the low-order model without topography for the same forcing heat flux $q^f$.}

\begin{figure}
\centering
\includegraphics[width=1\textwidth]{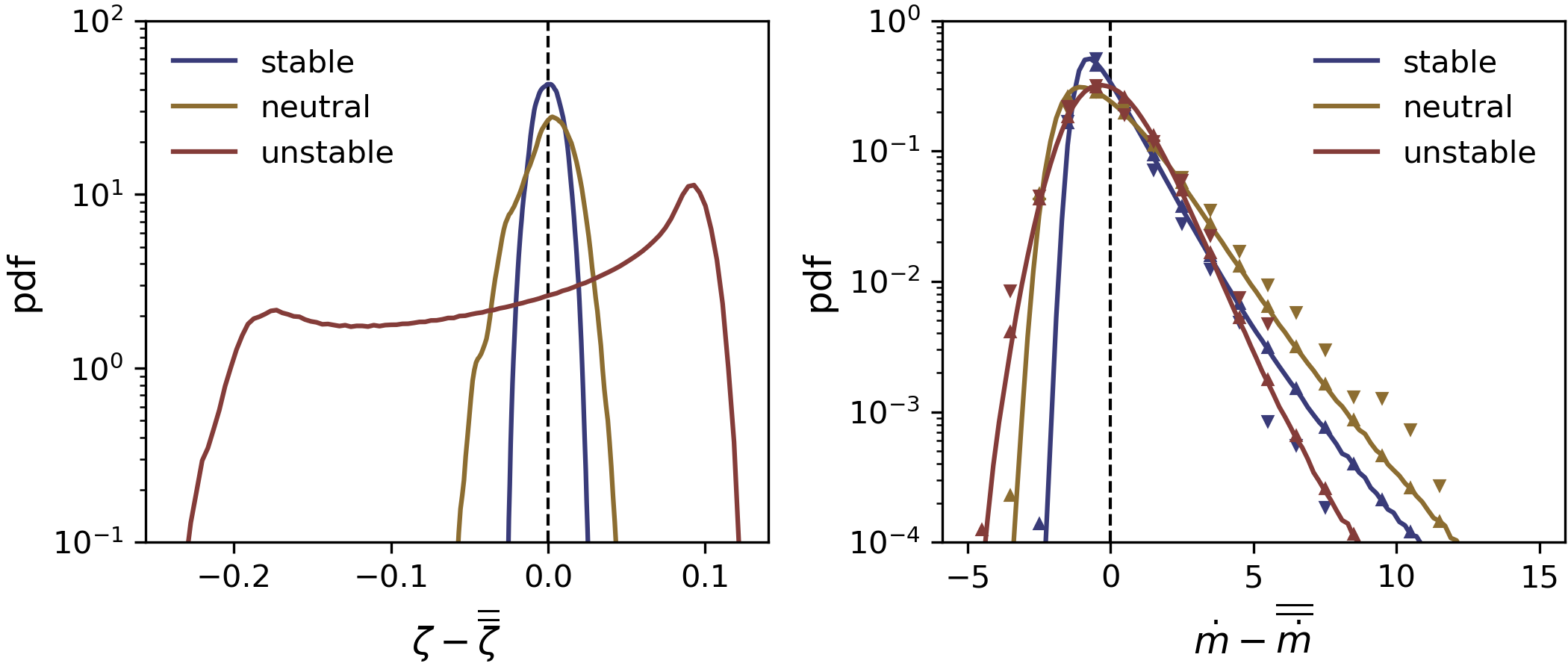}
\put(-385,160){\large{(a)}}\put(-190,160){\large{(b)}}
\vspace{-0.in}\caption{ Probability density functions (pdfs) of (a) the interface position $\xi$ and (b) the melt rate $\dot{m}$, for neutral, stable and unstable stratification. Double overlines denote temporal (100 friction time units) and spatial averaging. {In (b) we show the pdfs for the full melt rate (solid lines), the melt rate along channels (upper triangles) and the melt rate along keels (lower triangles).}}
\label{fig10}
\end{figure}

We finally show in figure \ref{fig10} the pdfs of interface deformation and melt rate {at statistical steady state, i.e., past the initial transient during which topographical features emerge and the solid melts on average. At statistical steady state, the pdfs of interface position become roughly time invariant. The mean interface position reaches a plateau (cf. figure \ref{fig9}(a)) because the mean amount of freezing balances the mean amount of melting at every time step, and the standard deviation, or topography amplitude, saturates (as can be seen in figure \ref{fig3}(f) for the unstable case). For stable stratification, the steady-state pdf of interface deformation  is almost symmetric with respect to the mean and appears approximately Gaussian (figure \ref{fig10}(a)). This suggests that channels and keels are symmetric of each other with respect to the mean interface position for the stable case. For neutral stratification, a small asymmetry develops, i.e., the median shifts toward small positive deformations (channels) and the tail of extreme negative deformations (keels) increases slightly. The same asymmetry is amplified for unstable stratification, with a narrow peak appearing to the right of the mean and the negative tail increasing further. In other words, as the stratification becomes unstable, patterns grow in size and the width-to-height ratio of channels increases (broad and flat) while the width-to-height ratio of keels decreases (narrow and deep). The asymmetry in the pdfs of interface topography is consistent with the observation from  figure \ref{fig3} that channels are typically flatter and more widespread than keels.} 

The pdfs for the melt rate $\dot{m}$ are shown in figure \ref{fig10}(b). The temporal and spatial average, $\overline{\overline{\dot{m}}}$, which is subtracted from the pdf, is close to 0 in all cases, since the mean amount of melting is balanced by the mean amount of freezing at statistical steady state. The pdfs of melt rate are asymmetric{, i.e., similar to the pdfs of heat flux at the top of fluid-only channel simulations (see dashed lines in figures \ref{fig7}(c),(f),(i))}. The median is shifted to the left of the mean, i.e., toward negative values representative of freezing events, and the positive tail is enhanced compared to the negative tail. In other words, the interface is typically freezing slowly ($\dot{m}<0$), but occasionally melts rapidly ($\dot{m}>0$). {Importantly, the asymmetry of the melt rate pdfs is not due to the asymmetry of the interface pdfs since the pdfs of melt rates inside channels (upper triangle) and along keels (lower triangles) are similar, but is instead a generic feature of melting by a turbulent fluid. Indeed, while the turbulent flow can drive rapid melting independently from what happens in the solid, freezing necessarily involves slow diffusive processes in the solid. Additionally, the near-wall dynamics, which features coherent structures such as streamwise streaks and vortices, is itself asymmetric, as can be seen from, e.g., the pdfs of temperature gradient in figure \ref{fig7}. Thus, it is not surprising that the melt rate pdfs are asymmetric. While beyond our goal, it would be worthwhile in the future to try identify  flow features controlling the shape of the melt rate pdfs.}

%

\section{Geophysical discussion}\label{sec:geophysics}

Due to the large computational costs of coupled fluid-solid simulations, all control parameters in this study were held fixed, i.e., we considered $Re_*=150$, $Pr=1$ and $St=1$, except for the friction Richardson numbers, which we varied in order to test the effect of density stratification on topography generation. {Our simulation with positive (resp. negative) Richardson number $Ri_*=40$ (resp. $Ri_*=-40$), i.e., with stable (resp. unstable) stratification, assumes a negative (resp. positive) thermal expansion coefficient. Cold freshwater has a negative (resp. positive) thermal expansion coefficient at low (resp. high) pressure \cite[][]{Thoma2010a}. Thus, our stable simulation is qualitatively similar to the flow of freshwater below an ice cover at low pressure (as is the case in an ice-covered lake), whereas our unstable simulation is qualitatively similar to the flow of freshwater below an ice cover at high pressure (as is the case in a deep subglacial lake). In the case where the solid is below the fluid, the stratification is reversed, i.e., such that the unstable simulation results are applicable to the flow of cold freshwater above ice at low pressure (as is the case in supraglacial rivers). Our stable simulation is also qualitatively similar to the flow of salt water under ice shelves. The melt water under ice shelves is cooler but also fresher than the ambient ocean water, such that it is positively buoyant. In fact, salinity and temperature can be combined, assuming that they have the same effective diffusivities, into a single variable known as thermal driving, which has a negative expansion coefficient \cite[][]{Jenkins2016}.} 

{In order to minimize resolution requirements and observe large topographical changes in a relatively small amount of time we have considered a flow that is only moderately turbulent, weakly stratified and anomalously warm. If we assume that the working fluid is water, i.e., such that $\mathcal{L}=3\times 10^5$ J/kg, $c_p=4\times 10^3$ J/kg/K, then $St=1$ implies that the bottom boundary is held at $75^{\circ}\mathrm{C}$. If we further assume that $H=10$ cm and $\nu=\kappa=10^{-6}$ m$^2$/s, i.e., considering an anomalously-high thermal diffusivity such that $Pr=1$, then our simulation runs correspond to about 10 hours in real time, the bulk velocity is 2 cm/s and the thermal expansion coefficient is $5\times  10^{-7}$ K$^{-1}$ in absolute value, which is actually too small for water. These calculations highlight that there is clearly a significant gap between the control parameters in our simulations and in nature. Our set of experiments is also limited to three different runs, such that we cannot offer a quantitative prediction of what would be observed in the environment. Nevertheless, Rayleigh-B\'enard rolls have been observed for a wide range of Rayleigh numbers and Reynolds numbers in mixed-convection simulations with unstable stratification \cite[][]{Pirozzoli2017}, such that large-scale channels and keels driven by these rolls may be expected whenever the stratification is unstable, i.e., e.g., in supraglacial rivers or deep subglacial lakes, especially when the external flow is weak.  The transverse wavelength of channels and keels maintained by global rolls is expected to be of the same order as the fluid depth, as is the case in our unstable simulation  (cf. figure \ref{fig8}(a), in which each wavelength accommodates two counter-rotating rolls with diameter equal to the mean fluid depth). We note that our unstable simulation has low $Ra=4.5\times 10^5$ and $Re_*=150$ compared to state-of-the-art simulations of mixed convection \cite[][]{Pirozzoli2017,Blass2020}, but more importantly has a relatively small bulk Richardson number $Ri_b=RaPr/Re_b^2\approx 0.1$ (defined as positive for an unstable stratification) and large Monin-Obukhov length $L_{MO}\approx 1$ (normalized by $H$). Thus, channels and keels can be expected for a broad range of bulk Richardson numbers of unstably stratified shear flows, i.e., $Ri_b\geq 0.1$, and to grow in size as $Ri_b$ increases. In the limit $Ri_b \rightarrow \infty$, channels and keels may be expected to eventually disappear. Indeed, as $Ri_b \rightarrow \infty$, the mean flow vanishes and buoyancy effects dominate, such that three-dimensional domes and cusps should emerge in place of streamwise features \cite[][]{Rabbanipour2018}.} Interestingly, large channels have also been observed at the base of ice shelves. However, these channels are unlikely to originate from Rayleigh-B\'enard rolls but rather from transverse perturbations of, e.g., the subglacial discharge or ice thickness at the grounding line {\cite[][]{Dallaston2015}}, since the stratification is in this case stable.

{For stable and neutral stratifications, we also observed channels and keels. Channels and keels with stable stratification ($Ri_*\geq 0$) are carved by boundary-attached momentum streaks rather than by global modes, however, such that their transverse wavelength is shorter than for the case of unstable stratification (although here the difference is weak given the small $Re_*$), and their amplitude is either equal or smaller than the viscous sublayer thickness, i.e., small. The shape and size of the small channels and keels obtained for  $Ri_*\geq 0$ are in stark contrast with the three-dimensional scallops, which have been observed in stably-stratified polar oceans and neutral laboratory experiments.} Scallops observed in the field and investigated in laboratory experiments have amplitudes on the order of few centimeters and wavelengths on the order of few tens of centimeters, i.e., such that they are tall and wide features compared to the viscous sublayer thickness, which is typically smaller than 1 mm in nature \cite[][]{Bushuk2019}. Previous experimental and theoretical works have found that the friction Reynolds number based on the scallop wavelength $\lambda$ usually satisfies $Re_*^{\lambda}=Re_*\lambda/H \geq O(1000-10000)$ and have always reported a scallop wavelength smaller than the fluid depth, i.e., $\lambda < H$ \cite[][]{Blumberg1974,Thomas1979,Claudin2017}. Considering the upper limit $\lambda = H$ means that scallops are predicted to emerge for $Re_*\geq O(1000-10000)$ in water, which is at least one order of magnitude higher than what we selected for our study and difficult to achieve numerically. Note, however, that the minimum $Re_*$ leading to scallops formation may be different for a fluid with control parameters $Pr=1$ and $St=1$ (as is the case in this work) instead of $Pr\approx 10$ and $St\approx 75$, as is the case for water, {and may also vary with the stratification strength. }

\section{Concluding remarks}\label{sec:discussion}

We have shown that streamwise channels and keels spontaneously emerge as the dominant topographical features of a fluid-solid boundary when the flow is pressure-driven, turbulent and thermally-stratified with $Re_*=150$, $Pr=1$ and $St=1$. We have investigated the effect of the background density stratification and found that the amplitude of the channels and keels increases with decreasing stratification. For unstable stratification ($Ri_*=-40$), the channels and keels couple strongly with Rayleigh-B\'enard rolls, which are energised and locked within the interface deformation pattern. For neutral stratification, a similar correlation is obtained between the flow dynamics and the interface deformation pattern. {However, the full-depth rolls are replaced with smaller and weaker boundary-attached momentum streaks, which do not provide a clear locking mechanism, i.e., such that the topography drifts.} For neutral ($Ri_*=0$) and stable stratification  ($Ri_*=40$) the channels and keels saturate either because of the absence of a positive feedback between topography and momentum streaks or because stabilizing buoyancy forces inhibit channel growth. For unstable stratification, the saturation is due to the fact that we impose the temperature at the top boundary. With an imposed heat flux at the top, the entire solid would melt rapidly and entirely provided that the stratification is unstable (not shown), which means that the choice of boundary conditions at the top of the solid can be critical. {Note that the growth of the fluid layer for unstable stratification is due to the positive feedback that melting has on the effective Rayleigh number of the convective fluid.} As the solid melts, the effective Rayleigh number increases, leading to further melting, which is stopped only if diffusion in the solid can eventually balance the increasing heat flux in the fluid.

The analysis of the melt rate statistics indicates that there is an asymmetry in melting and freezing, which may be related to the different melting/freezing dynamics (freezing relying primarily on slow diffusive processes in the solid) but also asymmetries in the flow statistics. Specifically, melting is highly localized and intense while freezing is widespread but weak. While beyond the scope of this study, it would be useful to identify whether coherent features of the near-wall turbulent flow, such as streamwise streaks and vortices, correlate preferentially with either melting or freezing events. 

The drag coefficient changes significantly depending on the type of stratification but is only weakly affected by the generation of topographical features, which is not unexpected in our case since streamwise channels and keels are smooth in the direction of the flow. Capturing three-dimensional topographical features, such as scallops, which do affect momentum and heat transfers \cite[][]{Bushuk2019}, in coupled fluid-solid simulations, would be a major achievement, which could complement fluid-only simulations at ice boundaries \cite[][]{Gayen2016,Keitzl2016,Keitzl2016b, Mondal2019, Vreugdenhil2019}. However, as discussed in section \S\ref{sec:geophysics}, scallops may require much higher Reynolds numbers to form than $Re_*=150$. In fact, the minimum $Re_*$ for scallops could be too high for a phase-field method on most supercomputers. {The cheapest test for evaluating the minimum Reynolds number leading to scallops would be to start the simulations with a longitudinally-wavy boundary and investigate the initial evolution. The runtime would be reduced to a minimum. However, a high resolution (higher than say $1024^3$ with a spectral code) would still be required. It is noteworthy that simulations of a pure fluid at a fixed wavy boundary would already be useful in helping verify or refine the most recent theoretical predictions on scallops formation and saturation \cite[][]{Claudin2017} and estimate the effect of the stratification strength, which has not yet been considered.} We remark that capturing scallops in a water environment would not only require higher $Re_*$ but also higher $St$ and $Pr$, which would both incur significant computational overhead. Higher $Pr$ results in thinner thermal boundary layers, which could impact the near-wall dynamics and, e.g., the asymmetry between melting and freezing. Higher $St$ results in slower melt rates, which could significantly change how interface patterns couple with transient flow features. In the case of unstable stratification, we might still expect that Rayleigh-B\'enard rolls couple with the interface deformation pattern for high $St$, since they are relatively stationary flow features at least in the strong shear regime \cite[][]{Pirozzoli2017}. For neutral stratification, however, the interface evolves over time scales similar to those of the flow dynamics for $St=1$ (figure \ref{fig3}(e)), such that increasing $St$ might significantly decrease the sensitivity of the interface topography to fluid anomalies. {Finally, freshening effects, which are critical to ice-ocean interactions, would require adding slowly-diffusing salt to the simulations, which constitutes yet another significant challenge for multi-phase DNS.}

{From a fundamental physics viewpoint, it would be interesting to investigate in details how topographical features are modified when phase changes are driven by dissolution rather than by melting. The fluid-solid boundary conditions (Stefan condition) and scalar diffusivities are different between dissolution and melting experiments. However, similar longitudinal and rippled patterns have been observed in both cases \cite[e.g.,][for dissolution]{Allen1971}. It would be also worthwhile to explore the effect of phase changes and topographical features on the onset of global modes and the large-scale organization of mixed-convection flows, which are of interests to many fields of physics and engineering \cite[][]{Kelly1994,Pabiou2005,Blass2020}.} Finally, it would be useful to investigate potential analogies between ice patterns due to melting and freezing and the formation of sand ripples and dunes, which have and continue to be extensively studied \cite[e.g.,][]{Charru2013,DuPont2014}.

\section*{Acknowledgements}

We gratefully acknowledge fruitful discussions with members of the ISOBL group at the British Antarctic Survey and the University of Cambridge. This project has received funding from the European Union's Horizon 2020 research and innovation programme under the Marie Sklodowska-Curie grant agreement 793450. We acknowledge PRACE for awarding us access to Marconi at CINECA, Italy.  

\section*{Declaration of Interests}

The authors report no conflict of interest.

\appendix

\section{Phase-field method}\label{sec:appA}

The phase-field method transforms the discontinuous two-phase two-domain problem into a continuous two-phase one-domain problem, which can be solved numerically using a pre-existing fluid code. In order to reproduce the original problem correctly, the resolution and parameters $A$, $B$, $C$ and  $\Gamma$ of the phase-field equations \eqref{03} must satisfy several constraints {\cite[][]{Hester2020}}, which can be verified a posteriori by diagnosing the flow properties in the solid and fluid phases. Here, for our choice of resolution, we have $A=6/(5St)$, $B=(16/\delta^2)\times 6/(5St)$ and $C=1$, with $\delta$ chosen such that it is equal to 2 times the local grid size at $z=1$ (initial interface position). Also, $\Gamma=(\delta/2.648228)^2$ and we require time steps always smaller than $\Gamma/2$.  

\begin{figure}
\hspace{0.4in}
\includegraphics[width=0.8\textwidth]{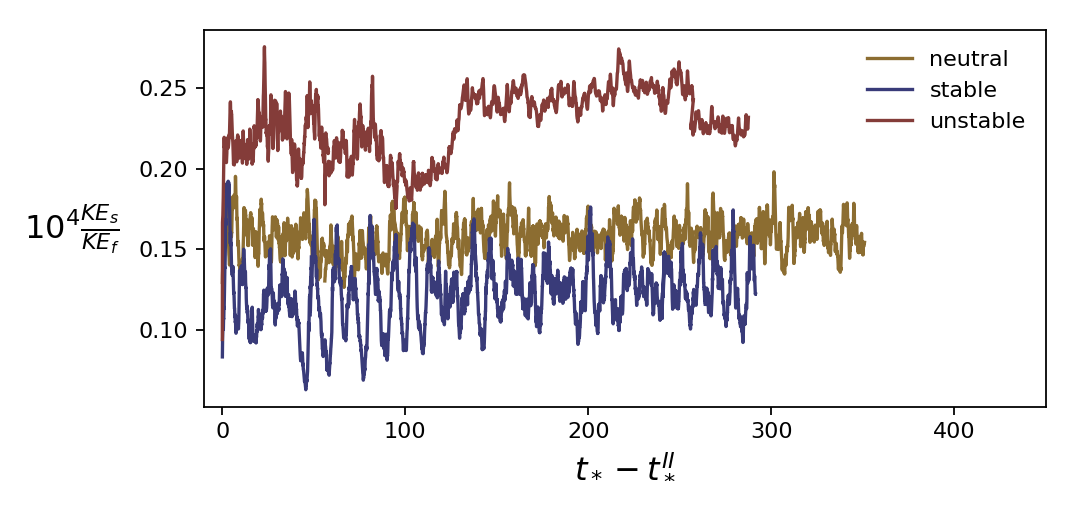}
\vspace{-0.1in}\caption{Kinetic energy in the solid divided by kinetic energy in the fluid as a function of time $t_*$. Red, gold and blue colors denote results obtained for unstable, neutral and stable stratification, respectively.}
\label{figapp1}
\end{figure}

We first assess the effect of the phase-field method and choice of parameters on the flow variables by showing in figure \ref{figapp1} the ratio of the kinetic energy averaged over the solid volume, $KE_s$, divided by the kinetic energy averaged over the fluid volume, $KE_f$. Figure \ref{figapp1} shows that $KE_s/KE_f<10^{-4}$ {and that the fluctuations are within a factor 2 of the mean, i.e., such that velocities in the fluid penetrate only very weakly into the solid and do not burst significantly.}

\begin{figure}
\centering
\includegraphics[width=1\textwidth]{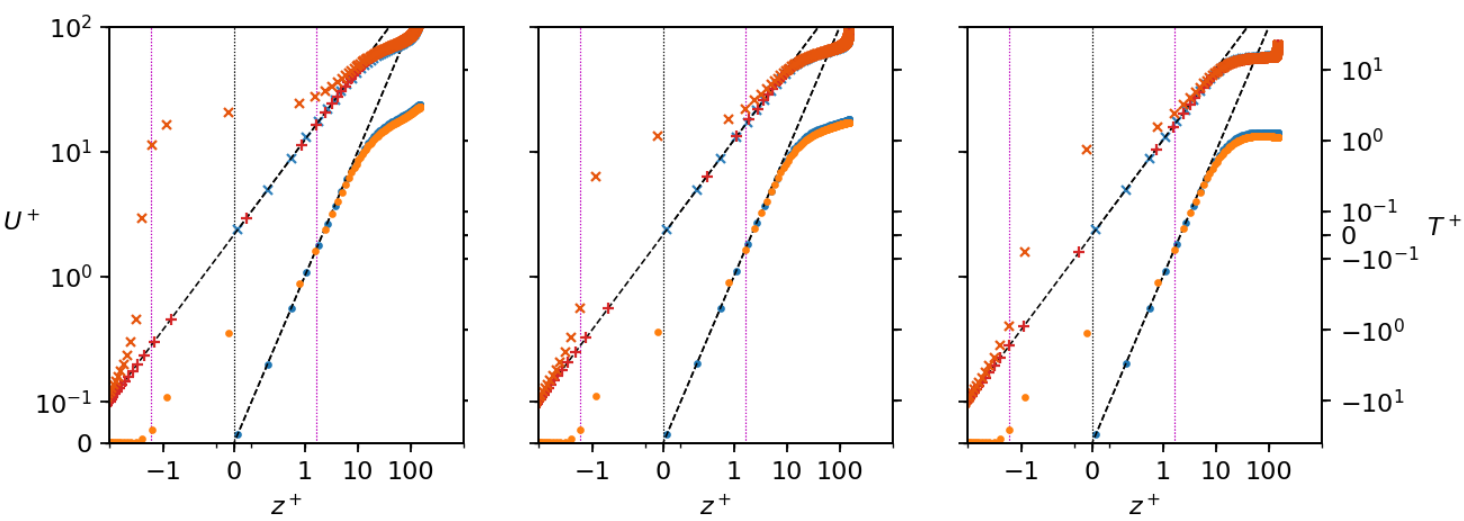}
\put(-385,130){\large{(a)}}\put(-260,130){\large{(b)}}\put(-147,130){\large{(c)}}
\vspace{-0.in}\caption{Wall-normalized velocity $U^+=\overline{Pru/u_*}$ (circles, left axes) and temperature $T^+=\overline{u_*T/(PrNu)}$ (crosses, right axes) expressed in terms of wall units as functions of wall coordinate $z^+=\overline{u_*(1-z)/Pr}$ for (a) stable, (b) neutral and (c) unstable stratification. Blue symbols denote results obtained for the simple channel flow configuration (i.e. for $t_*\in[t_*^{Ib},t_*^{Ic}]$) whereas orange symbols denote results obtained with volume penalization (i.e. for $t_*\in[t_*^{Ic},t_*^{II}]$). The plus symbols show the temperature profiles obtained with volume penalization but shifted to the right, i.e. into the fluid, by substituting $z^+$ with $z^++2.5$, $z^++1.16$ and $z^++0.78$ for stable, neutral  and unstable stratification. The overbars denote horizontal averaging and time averaging over 20 friction time units. The vertical dotted lines show $z^+=-\delta,0,\delta$ where $\delta$ is the diffuse interface thickness.}
\label{figapp2}
\end{figure}

{We now further comment on the resolution requirements and our choice for the grid size and time step.} We recall that $\delta$ is the thickness of the diffuse phase-field interface over which $\phi$ transitions from 1 in the fluid to 0 in the solid (see figure \ref{fig1}(c)). $\delta$ is an artificial length scale, such that it musts be smaller than any physical length scale in the problem, while at the same time being larger than the grid size since it must be resolved numerically. {For a boundary layer flow with $Pr=1$, the smallest length scale close to the interface is the viscous sublayer thickness, which is typically equal to a few times the viscous length scale $\delta_{\nu}=1/Re_*$. Here, we chose to have $\delta\approx\delta_{\nu}$ in all simulations, i.e., $\delta$ is equal to one wall unit $\Delta z^+=\delta$, such that the diffusive interface for the phase-field is comprised within the viscous and thermal sublayers, as can be seen in figure \ref{figapp2}. In order to resolve the diffusive interface, we selected a vertical Chebyshev basis with enough modes such that the collocation grid has a resolution $dz$ near the mean interface position equal to or less than $\delta_{\nu}/2$. The damping time scale $\Gamma$ in equation \eqref{03} is set to $\Gamma=(\delta/2.648228)^2$ in order to cancel first-order errors in the phase-field model \cite[][]{Hester2020}.}

Figures \ref{figapp2}(a-c) show a semilog plot of the wall-normalized velocity $U^+$ (left axes) and wall-normalized temperature $T^+$ (right axes) as functions of wall units $z^+$ ($z^+\geq 0$ denote positions in the fluid while $z^+<0$ denote positions in the solid) in stages I and II for stable, neutral and unstable stratification, respectively. In the viscous and thermal sublayers, which extend from $z^+=0$ to $z^+\approx 5$, we expect a linear scaling for both $U^+$ and $T^+$ with $z^+$, shown by the solid dashed lines. This linear scaling is perfectly satisfied by the DNS results in stage I (blue circles and blue crosses) as well as the DNS results in stage II (orange circles and crosses), except for $|z^+|<\delta$ (shown by the vertical solid lines), i.e. within the diffuse interface, which is expected since this is where the dynamics is artificially controlled by the phase-field equation. It may be noted that $T^+$ (orange crosses) is anomalously large for $z^+<\delta$ and in fact deviates from the true solution (blue crosses) slightly outside the diffuse interface. This discrepancy is due to the fact that the heat flux in the fluid is larger than the heat flux in the solid in stage II. The interface being fixed in stage II, the heat imbalance results in the heating of the solid, such that $T^+=0$ occurs at $z^+<0$ away from the fixed interface position $z^+=0$ (note that we use a symmetric logarithmic scale with a linear threshold at $|z^+|=0.1$). By shifting the temperature profile to the right such that $T^+=0$ is aligned with $z^+=0$ (red pluses), we recover a perfect linear scaling for the temperature both within the thermal fluid sublayer and the solid. {Outside of the linear sublayer and the buffer layer, the mean vertical profiles exhibit a logarithmic behavior.} 

Far from the top boundary, i.e., e.g., for $z^+\approx 100$, $U^+$ shows a steeper scaling with $z^+$ for stable stratification than for neutral of unstable stratification. This is a consequence of buoyancy effects, {which tend to decrease (resp. increase) stirring of the mean flow when the stratification is stable (resp. unstable)} {} \cite[][]{Garcia-Villalba2011}.

\section{Depth-independent variables}\label{sec:appB}

At statistical steady state, stratified pressure-driven flows between solid boundaries have linearly-varying shear stress $\overline{\tau}=\overline{\tau_{\nu}+\tau_w}$, with $\tau_{\nu}=\overline{\p_z u}$ the viscous stress and $\tau_w=-\overline{wu}$ the Reynolds stress, and depth-independent heat flux $\overline{q}=\overline{wT-\p_zT}$, where overbar denotes horizontal and time averaging. These conservation equations for the vertical fluxes of momentum and heat are at the origin of the definitions of the friction velocity and Nusselt number, which typically read  $u_*=\sqrt{-\overline{(\tau_{\nu}+\tau_w)}}|_{z=1}$ and $Nu=\int_0^{1} \overline{q} dz$ (assuming $z=1$ is the top of the fluid), respectively. With the phase-field method, these conservation equations are modified and some of the modifications are reflected in the definitions of $u_*$ and $Nu$ in equation \eqref{definitions}. In particular, $u_*$ in equation \eqref{definitions} includes the linear damping term $\tau_d=-\int_0^{z} (1-\phi)u/\Gamma dz'/z$ that comes from the last term on the right-hand-side of equation \eqref{031}. The true conservation of vertical momentum and heat fluxes based on governing equations \eqref{03} read
\ba{}
\p_z \overline{\lp \tau + \tilde{\tau} \rp} = -2Pr^2Re, \quad \p_z \overline{\lp q+\tilde{q} \rp} = 0 ,  
\ea 
where $\tilde{\tau}$ and $\tilde{q}$ are the anomalous stress and heat flux due to damping of the advective terms in equations \eqref{03}, i.e.,
\ba{}
\tilde{\tau} = \int_0^{z} -(1-\phi)(\u\cdot\NA)u dz'/z,  \quad \tilde{q} = \int_0^{z} -(1-\phi)(\u\cdot\NA)T dz'/z.
\ea
The existence of anomalous stress and heat fluxes means that the friction velocity and Nusselt numbers as defined in equation \eqref{definitions} are based on a total stress and heat flux, which are not rigorously linearly varying or depth invariant.

\begin{figure}
\centering
\includegraphics[width=1\textwidth]{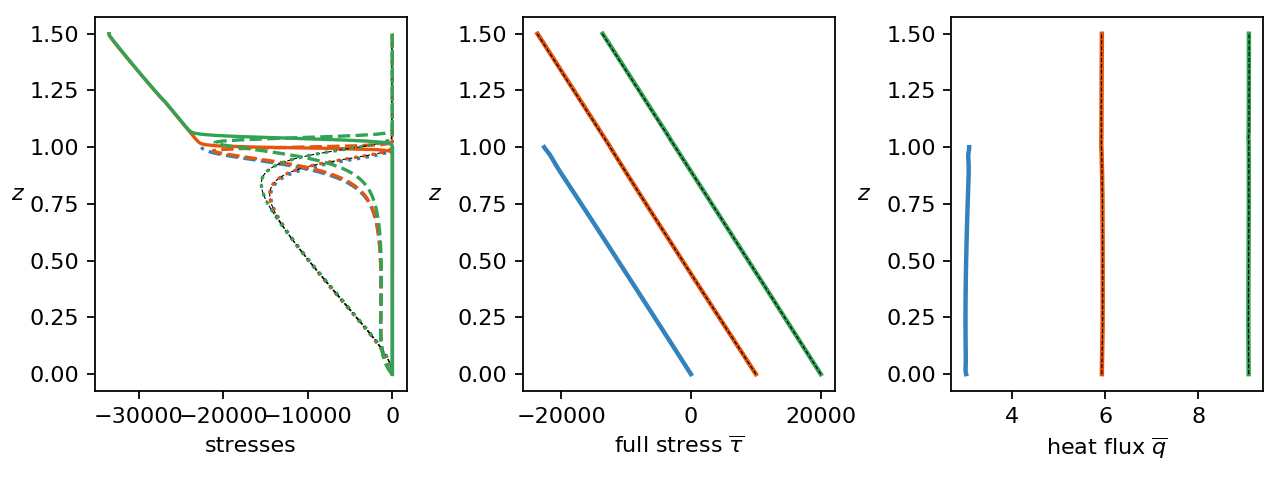}
\put(-370,145){\large{(a)}}\put(-242,145){\large{(d)}}\put(-113,145){\large{(g)}}\\
\includegraphics[width=1\textwidth]{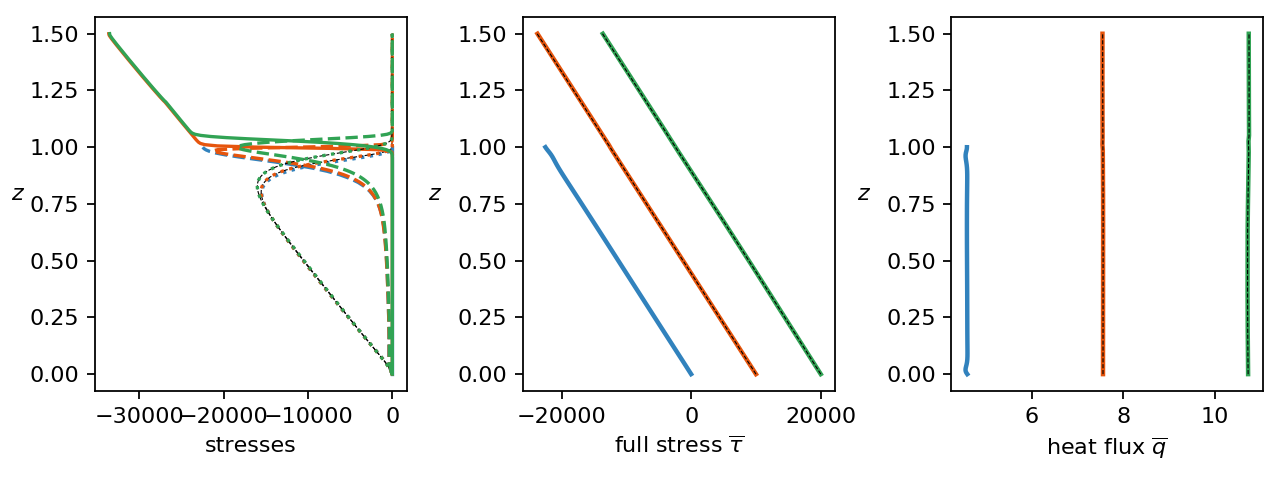}
\put(-370,145){\large{(b)}}\put(-242,145){\large{(e)}}\put(-113,145){\large{(h)}}\\
\includegraphics[width=1\textwidth]{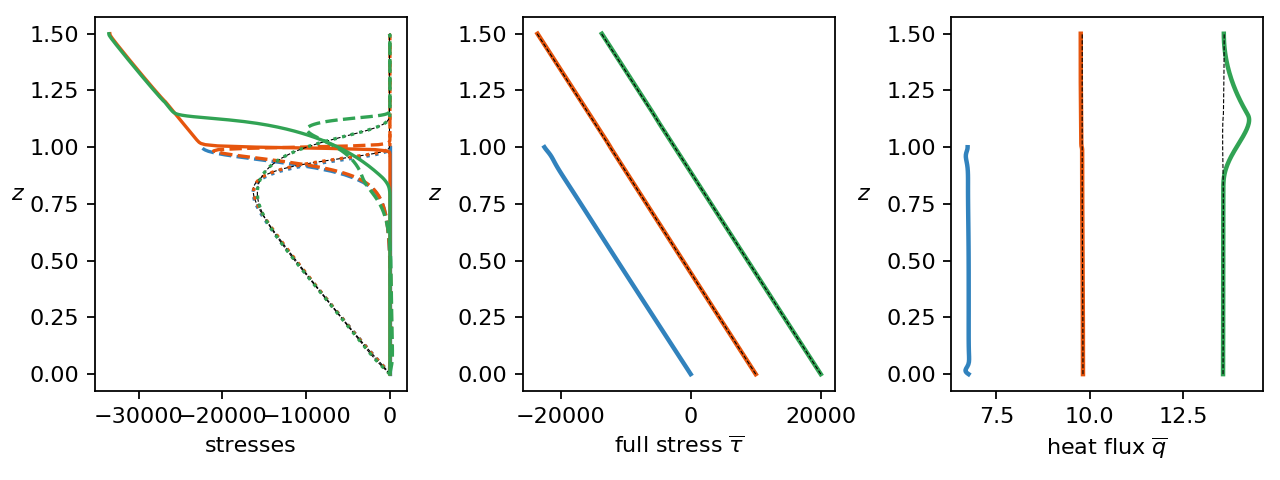}
\put(-370,145){\large{(c)}}\put(-242,145){\large{(f)}}\put(-113,145){\large{(i)}}
\vspace{-0.1in}\caption{(a-c) Reynolds stress $\tau_w$ (dotted lines), viscous stress $\tau_{\nu}$ (dashed lines), linear damping stress $\tau_d$ (solid lines) and Reynolds stress plus anomalous stress $\tau_w+\tilde{\tau}$ (for stages II and III; black dashed lines) averaged in time over 20 friction time units and horizontal planes as functions of depth $z$ for stable, neutral and unstable stratification, respectively (see the text for more details). Blue, orange and green colors denote results obtained in stage I, II and III, respectively (same in (d-f) and (g-i)). (d-f) Total stress, i.e.  $\tau_w+\tau_{\nu}+\tau_d$ (solid lines), and total stress plus the anomalous stress, i.e.  $\tau_w+\tau_{\nu}+\tau_d+\tilde{\tau}$ (thin dashed lines), averaged in time over 20 friction time units and horizontal planes as functions of depth $z$ for stable, neutral and unstable stratification, respectively. Note that the full stress is shifted to the right by 10000 (20000) between stage II and stage I and between stage III and stage II  for the case of stable (unstable) stratification for clarity as all curves overlap otherwise. (g-i) Heat flux $q$ (solid lines) and heat flux plus anomalous heat flux $q+\tilde{q}$  (thin dashed lines) averaged in time over 20 friction time units and horizontal planes as functions of depth $z$ for stable, neutral and unstable stratification, respectively. Note that the heat flux is shifted to the right by 2.5 (5) between stage II and stage I and between stage III and stage II for the case of stable (unstable) stratification  for clarity.}
\label{figapp3}
\end{figure}
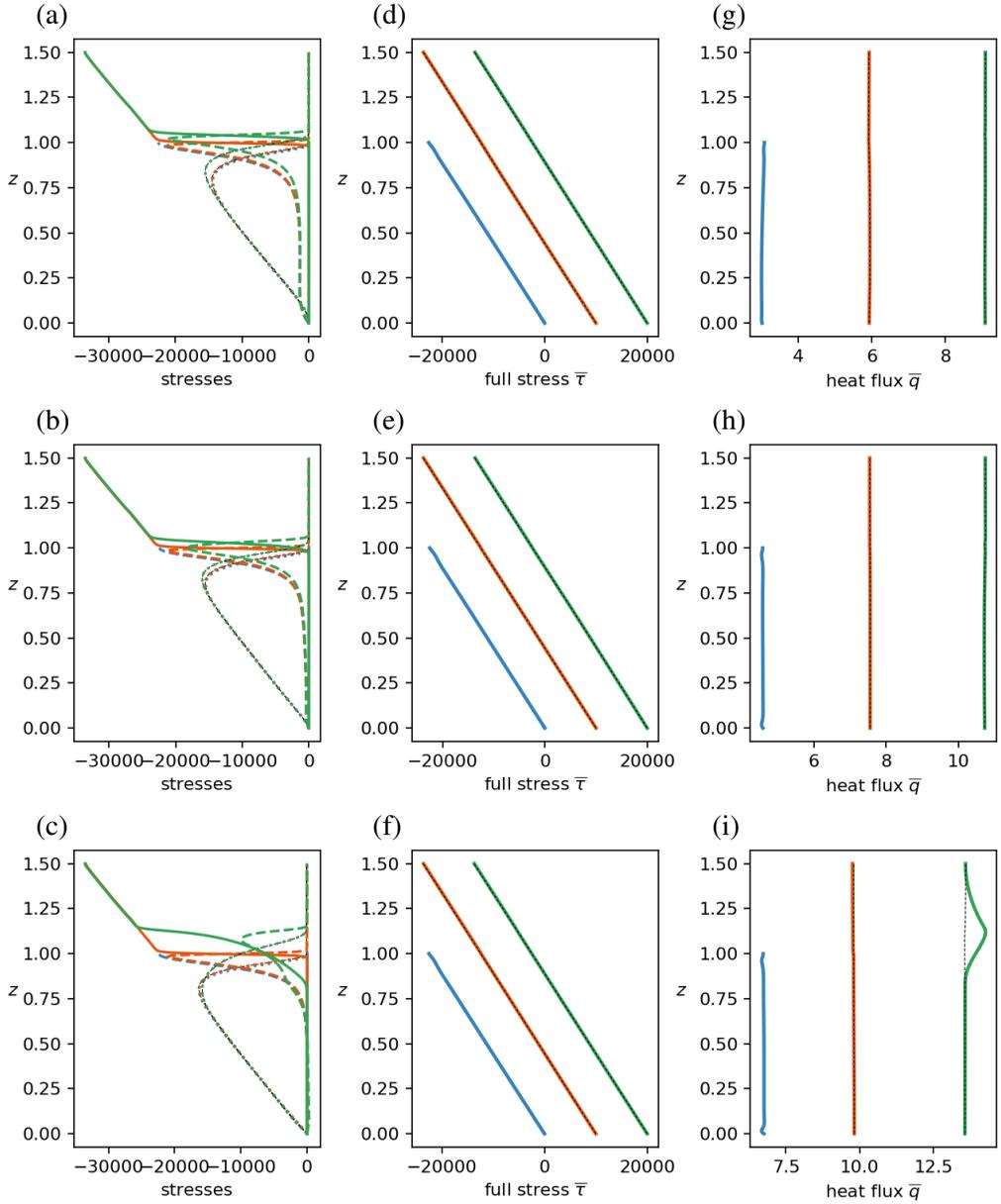

We show in figures \ref{figapp3}(a-c) the Reynolds stress $\overline{\tau_w}$, the viscous stress $\overline{\tau_{\nu}}$, the linear damping stress $\overline{\tau_d}$ and the Reynolds stress plus the anomalous stress $\overline{\tau_w+\tilde{\tau}}$ for stable, neutral and unstable stratification, respectively. Importantly, $\overline{\tau_w}$ and $\overline{\tau_w+\tilde{\tau}}$ overlap  well, showing that the anomalous stress is negligible. The results of figures \ref{figapp3}(d-f) further confirms that the anomalous stress is negligible in all simulations: the (approximate) total stress (solid lines) decreases linearly with $z$ in all stages and overlap well with $\overline{\tau_w+\tau_{\nu}+\tau_d+\tilde{\tau}}$, i.e., the total stress that includes the anomalous stress. We show in figures \ref{figapp3}(g-i) $\overline{q}$ (solid lines) and $\overline{q+\tilde{q}}$ (thin dashed lines). For stable and neutral stratification, $\overline{q}$  and $\overline{q+\tilde{q}}$ are constants with depth and overlap perfectly, suggesting that the anomalous heat flux is negligible. For unstable stratification, we obtain similar results for stages I and II. For stage III, however, $\overline{q}$ is not perfectly constant, but deviates from $\overline{q+\tilde{q}}$ and peaks at $z\approx 1.15$, which is roughly the height of the channels. The relative discrepancy between $\overline{q}$ and $\overline{q+\tilde{q}}$ is on the order of 5\% and is a result of the damping of the advective terms in the momentum and heat equations \eqref{03}. We expect that this discrepancy would decrease with increased resolution. Previous studies have alternatively considered advective terms with the same damped form as here, with a divergence damped form, i.e., $(\u\cdot\NA)(\phi\u)$, or without any damping. There is no proof that any of these methods is more efficient than the other two. However, we would recommend using either one of the latter two methods, i.e., not the method used in this paper, in order to simplify the analysis of the shear stress and heat flux.

\section{Additional details on the simulation stages}\label{sec:appC}

In this section we give additional details on the simulation stages and sub-stages. For $t\leq t_*^{Ic}$ (including stage I), we solve equations \eqref{031}-\eqref{032} and \eqref{034} with $\phi\equiv 1$ and a no-slip isothermal top boundary condition, i.e., {$\u=\bf{0}$} and $T=0$ at $z=1$. We use a straightforward half channel flow configuration, i.e., without a solid domain, with 64 Chebyshev modes in the vertical direction. In stage II we add a solid layer of thickness 0.5 on top of the fluid domain and we use a compound Chebyshev basis stitched at $z=1.2$ with 256 (resp. 32) Chebyshev modes in the lower (resp. upper) region. The compound Chebyshev basis allows to have a high vertical resolution near the interface's initial position. We solve equations \eqref{031}-\eqref{032} and \eqref{034} with $\phi$ prescribed, i.e., not varying in time (cf. main text). In stage III we solve equations \eqref{03} with all variables freely evolving and we use the same spectral resolution as in stage II.

Our simulations until $t = t_*^{Ic}$ can be broken down into three sub-stages. In stage Ia, i.e., for $t_*<t_*^{Ia}$ (cf. light gray colors in figure \ref{fig2}), we run a low-resolution (128 Fourier modes in $x$ and $y$ directions and 32 Chebyshev modes in $z$ direction) spin-up simulation of an initially-laminar flow superposed with three-dimensional velocity perturbations and no buoyancy effects ($Ri_*=0$). In stage Ib, i.e., for $t_*^{Ia}\leq t_* < t_*^{Ib}$, we increase the resolution (256 Fourier modes in $x$ and $y$ directions and 64 Chebyshev modes in $z$ direction) but keep $Ri_*=0$ (cf. dark gray in figure \ref{fig2}). In stage I, we turn on buoyancy effects ($t_*^{Ib}\leq t_* < t_*^{Ic}$; cf. blue in figure \ref{fig2}), i.e., we use 
\ba{}\label{05}
Ri_*(t) = Ri_*\tanh\lp t_*-t_*^{Ib} \rp, 
\ea
such that $Ri_*$ transitions smoothly (over the time scale of one friction time unit) from 0 at the end of stage Ib to the target value listed in table \ref{table1}. Note that in the case of stabilizing buoyancy effects, we found that the turbulent flow relaxes to a laminar state when using \eqref{05}. In order to avoid this we used an intermediate stage with a more moderate target $Ri_*=20$ (sub-stage Ici) before transitioning to $Ri_*=40$, using a similar equation as \eqref{05}.

\section{Higher-order interface evolution model}\label{sec:appD}

In this section we derive a reduced model for the evolution of the mean interface position $\overline{\xi}$, or ice thickness $h=h_0-(\overline{\xi}-{\xi}_0)$, with $h_0=1/2$ and ${\xi}_0=1$ the initial ice thickness and interface position, which takes into account interface deformation. Under steady-state assumption (instantaneous temperature diffusion), the evolution of $\overline{\xi}$ is controlled by the difference between the input heat flux from the fluid, $q^f$, and the mean heat flux at the ice top, $q_{top}$, i.e.
\ba{}\label{D1}
\f{d\overline{\xi}}{dt} = q^f-q_{top}.
\ea
At leading order, we assume that the interface is flat such that $q_{top}=h_0q^s/h$, which yields equation \eqref{toy}. At higher order, we take into account interface deformation, which can change $q_{top}$, using regular perturbation \cite[][]{Favier2019}. Specifically, we seek a solution of the $x$-averaged steady-state heat equation, i.e., (dropping tilde for $x$-averaged variables)
\ba{}
\nabla^2 T = 0, \quad &\xi(y) \leq z \leq L_z, \\
T = -h_0q^s, \quad &z=L_z, \\
T = 0, \quad &z=\xi(y), 
\ea
using a perturbation series of the form
\ba{}
T(y,z) = T^{(0)}(z) + T^{(1)}(y,z) + T^{(2)}(y,z) ..., 
\ea
with $\p_yT^{(0)}\equiv 0$ and $T^{(i)} \sim O(\epsilon^i)$, where $\epsilon\ll 1$ is the dimensionless amplitude of the interface deformation. For simplicity, here we approximate the interface deformation as
\ba{}
\xi(y) = \overline{\xi} + \epsilon \cos ky.
\ea
The leading-order solution is 
\ba{}
T^{(0)} = -\lp\f{z-\overline{\xi}}{L_z-\overline{\xi}}\rp h_0q^s,
\ea
the first-order solution is
\ba{}
T^{(1)} = \lp\f{\epsilon h_0q^s}{L_z-\overline{\xi}} \rp \cos kx \f{\sinh k\lp L_z-z \rp}{\sinh k\lp L_z-\overline{\xi}\rp},
\ea
and the second-order solution is
\ba{}
T^{(2)} = \f{\epsilon^2h_0q^s}{2\lp L_z-\overline{\xi}\rp \tanh k \lp L_z-\overline{\xi} \rp}  \lb \lp\f{L_z-z}{L_z-\overline{\xi}}\rp + \cos 2kx \f{\sinh 2k\lp L_z-z \rp}{\sinh 2k\lp L_z-\overline{\xi}\rp} \rb.
\ea
Thus, the second-order accurate formula for the mean heat flux at the top of the ice reads
\ba{}\label{D10}
q_{top} = -\overline{\p_z T}(z=L_z) \approx \f{h_0q^s}{h}\lb 1+ \f{\epsilon^2k}{2h\tanh kh} \rb,
\ea
which differs from the leading-order heat flux only at second order. The prediction for the evolution of the mean interface position for unstable stratification with $q^f=Nu^{III}$, which is shown by the red dashed lines in figure \ref{fig9}(a), is based on equation \eqref{D1} with $q_{top}$ given by \eqref{D10} and with $\epsilon=0.137$ and $k=2$ (as obtained from best-fit of the true interface topography at steady state for unstable stratification). We note that the quasi-steady state assumption may affect the prediction of the transient evolution of the mean interface position adversely but has no effect on the final value, which is the primary goal of the reduced model.

\bibliographystyle{jfm}
\bibliography{icemelting}

\end{document}